\documentclass[preprint,review,12pt]{elsarticle}
\usepackage{psfig,epsf,rotating,aalongtable,txfonts}
\usepackage{color}

\begin{document}
\begin{frontmatter}
\baselineskip=15pt
\textwidth 16.0truecm
\textheight 21.0truecm
\oddsidemargin 0.5truecm
\evensidemargin  0.3truecm
\topmargin 0.2cm
\headsep 1.0cm

\title{Aqueous alteration on main belt primitive asteroids: results from visible spectroscopy\footnote{Based on observations carried out at the European Southern
Observatory (ESO), La Silla, Chile, ESO proposals 062.S-0173 and 064.S-0205 (PI M. Lazzarin)}}
\author{S. Fornasier$^{1,2}$, C. Lantz$^{1,2}$, M.A. Barucci$^{1}$, M. Lazzarin$^{3}$}

\maketitle
\noindent
$^1$ LESIA, Observatoire de Paris, CNRS, UPMC Univ Paris 06, Univ. Paris Diderot,
5 Place J. Janssen, 92195 Meudon Pricipal Cedex, France \\
$^2$ Univ. Paris Diderot, Sorbonne Paris Cit\'{e}, 4 rue Elsa Morante, 75205 Paris 
Cedex 13 \\
$^3$ Department of Physics and Astronomy of the University of Padova, Via Marzolo 8 35131 Padova, Italy\\
\noindent
Submitted to Icarus: November 2013, accepted on 28 January 2014 \\
e-mail: sonia.fornasier@obspm.fr; fax: +33145077144; phone: +33145077746\\
Manuscript pages: 38; Figures: 13 ; Tables: 5  \\
\vspace{3cm}

{\bf Running head}: Aqueous alteration on primitive asteroids

\noindent

{\it Send correspondence to:}\\
Sonia Fornasier  \\
LESIA-Observatoire de Paris  \\
Batiment 17 \\
5, Place Jules Janssen \\
92195 Meudon Cedex \\
France\\
e-mail: sonia.fornasier@obspm.fr\\
fax: +33145077144\\
phone: +33145077746\\

\newpage
\vspace{2.5cm}

\begin{abstract}

This work focuses on the study of the aqueous alteration process which acted in the main belt and produced hydrated minerals on the altered asteroids. Hydrated minerals have been found mainly on Mars surface, on main belt primitive asteroids and possibly also on few TNOs. These materials have been produced by hydration of pristine anhydrous silicates during the aqueous alteration process, that, to be active, needed the presence of liquid water
under low temperature conditions (below 320 K) to chemically
alter the minerals. The aqueous alteration is particularly important for unraveling the processes occurring during the earliest times of the Solar System history, as it can give information both on the asteroids thermal evolution and on the localization of water sources in the asteroid belt.  \\
To investigate this process, we present reflected light spectral observations in the visible region (0.4--0.94 $\mu$m) 
of 80 asteroids belonging to the primitive classes C (prevalently), G, F, B and P, following the
Tholen (1984) classification scheme. 
We find that about 65 \% of the C-type  and all the G-type asteroids investigated reveal features suggesting the presence of hydrous materials, mainly a band centered around 0.7 $\mu$m, while we do
not find evidence of hydrated materials in the other low albedo
asteroids (B, F, and P) investigated. \\
We combine the present observations with the visible spectra of asteroids available in the literature for a total of 600  primitive main belt asteroids.  We analyze all these spectra in a similar way to characterize the absorption band parameters (band center, depth and width) and spectral slope, and to look for possible correlations between the aqueous alteration process and the asteroids taxonomic classes, orbital elements, heliocentric distances, albedo and sizes. 
Our analysis shows that the aqueous alteration sequence starts from the  P-type objects, practically unaltered, and increases through the P $\rightarrow$ F $\rightarrow$ B $\rightarrow$ C $\rightarrow$ G asteroids, these last being widely aqueous altered, strengthening thus the results previously obtained by Vilas (1994). Around 50\% of the observed C-type asteroids show absorption feature in the visible range due to hydrated silicates, implying that more than $\sim$ 70\% of them will have a 3 $\mu$m absorption band and thus hydrated minerals on their surfaces, based on correlations between those two absorptions (Howell et al., 2011). \\
We find that the aqueous alteration process dominates in primitive asteroids located between 2.3 and 3.1 AU, that is at smaller heliocentric distances than previously suggested by Vilas et al. (1993). The percentage of hydrated asteroids is strongly correlated with their size. The aqueous alteration process is less effective for bodies smaller than 50 km, while it dominates in the 50--240 km sized primitive asteroids.  \\
No correlation is found between the aqueous alteration process and the asteroids albedo or orbital elements. Comparing the $\sim$ 0.7 $\mu$m band parameters of hydrated silicates and CM2 carbonaceous chondrites, the meteorites that have aqueous altered asteroids as parent bodies,  we see that the band center of meteorites is at longer wavelengths than that of asteroids. This difference on center positions may be attributed to different minerals abundances, and to the fact that CM2 available on Earth might not be representative of the whole aqueous altered asteroids population.

\end{abstract}

\end{frontmatter}

\newpage
\section{Introduction}

The distribution of the asteroids taxonomic classes follows a radial structure varying with the  heliocentric distance: the inner and middle belt is dominated by differentiated bright asteroids (M, E, V, and prevalently S-types) that have experienced high temperatures in the past and that are constituted by volatile-poor silicates and metals, while dark primitive asteroids (B, P, D, and prevalently C-types), rich in carbonaceous chondrite-like materials and formed in a colder environment, dominate the outer belt to the Jupiter Trojans. This asteroid class distribution corresponds to a radial variation of the formation temperatures and to the presence of different mineralogical materials with increasing heliocentric distance. These materials have also been partially mixed in the past due to dynamical interaction of asteroids with each others and with planets, and altered over time by different processes such as collisions, aqueous alteration, and space weathering.

The aqueous alteration process acts on primitive asteroids (C, G, B, F and P-types, following the Tholen (1984) classification scheme) located
mainly between 2.6 and 3.5 AU, in the so called zone of aqueous alteration (Vilas, 1994; Fornasier et al., 1999). The aqueous alteration process produces the low temperature ($<$ 320 K) chemical alteration of materials by liquid water which acts as a solvent and produces hydrated minerals such as phyllosilicates, sulfates, oxides, carbonates, and hydroxides. This means that liquid water was present in the primordial asteroids, produced by the melting of water ice by heating sources, very probably by  $^{26}$Al decay. \\
Reflectance spectroscopy of aqueous altered asteroids shows absorption features in the 0.6-0.9 and 2.5-3.5 micrometers regions, which are diagnostic of, or associated with, hydrated minerals.  The most prominent and unambiguous indicator of hydratation is the 3 $\mu$m band (Lebofsky, 1980; Jones et al.,
1990; Rivkin et al., 2002; Vilas, 1994, 2005; Howell et al., 2011; Takir \& Emery, 2012),  associated with {\it free} water
molecules, and to OH ion bounded in
the mineral crystal lattice; while in the visible range there are several
bands, centered around 0.43 $\mu$m, 0.60--0.65 $\mu$m, 0.70 $\mu$m and
0.80--0.90 $\mu$m, attributed to charge transfer transitions in oxidized 
iron (Vilas et al., 1993, 1994; Vilas, 1994; Barucci et al., 1998; Fornasier et al., 1999).

The study of aqueous alteration is particularly important for unraveling the processes occurring during the
earliest times in Solar System history, as it can give information both on the thermal processes and on the localization of water sources in the asteroid belt, and for the associated astrobiological implications.  Indeed it has been suggested (Morbidelli et al., 2000; Lunine, 2006) that the Earth current supply of water was delivered mostly by asteroids, not comets, some time following the collision that produced the Moon (which would have vaporized any of the water existing at that time). While asteroids and comets from the Jupiter-Saturn region were deemed to be the first water deliverers (when the Earth was half its size), the bulk of the water delivered was found to originate from a few planetary embryos,
originally formed in the outer asteroid belt and accreted by the Earth at the final stage of its formation. Morbidelli et al. (2000) found that the measured amount of the D/H ratio in the water on Earth correlates very well with the ratio of D/H typical of water
condensed in the outer asteroid belt. Alexander et al. (2012), analyzing the bulk hydrogen and nitrogen isotopic compositions of CI chondrites, suggest that these meteorites were the principal source of Earth  volatiles, while the D/H of all comets except Hartley 2 (Hartogh et al., 2012) are quite different from the Earth oceans value. 

The snow line has been located within or nearby the asteroid belt (Lunine, 2006; Cyr et al., 1998), and, following the recent nebular kinetics models results, it may
have migrated as the nebula evolved, sweeping across the entire asteroid belt (Dodson-Robinson et al., 2009). According to the latest dynamical models, it has also been suggested that part of the main belt primitive asteroids were formed either between the giant planets or in the Trans-neptunian region (Walsh et al., 2011), and then injected into the inner Solar System. \\
Water ice has been tentatively identified on the surfaces of the outer primitive asteroids Themis and Cybele (Rivkin and Emery, 2010; Campins et al., 2010; Licandro et al., 2011), and  some main belt asteroids, originated from the Themis family (Hsieh \& Jewitt, 2006; Hsieh et al., 2009, 2012), were discovered to become active (they are also called main belt comets).  

The fact that primitive asteroids had retained water ice in the past and may have enriched the Earth of water and possibly organic material, favoring the appearance of life on our planet, is of great interest for the scientific community. Indeed all of the major space agencies are planning sample return missions to primitive near Earth objects (NEO): NASA will launch OSIRIS-REx in 2016 to sample the asteroid (101955) Bennu, JAXA will launch Hayabusa-2 in 2014 to bring back material from (162173) 1999 JU3, and the mission MarcoPolo-R has been proposed to the European Space Agency (ESA) in the framework of the Cosmic Vision 2015--2025 program, for a sample return from (342843) 2008 EV5. \\
Some evidence of hydrated material has been found also in the outer Solar System, with the detection of peculiar absorption bands on the spectra of some TNOs that could be possibly  associated with the aqueous
alteration process (Fornasier et al., 2004a, 2009; Lazzarin et al., 2003; de Bergh et al., 2004; Alvarez-Candal, 2008).

In this work we focus on spectroscopy in the visible region of low albedo asteroids. We present new visible spectra for 80 objects belonging to the C-complex and localized between 2.3 and 4.0 AU, with the aim to investigate the aqueous alteration process that has involved the  C, B, F, P, and G primitive classes. To do so, we have studied a larger sample including visible spectra from the literature for a total of 600 primitive main belt asteroids. We have analyzed all these data to spectrally characterize the bands  associated with the aqueous alteration process in the visible region (mainly the one centered around 0.7 $\mu$m), and look for the relationships between
this process and the heliocentric distance, albedo and diameter of the investigated objects. We have also compared the spectral parameters  associated with the 0.7 $\mu$m band for asteroids and for the CM2 carbonaceous chondrites, which show evidence of aqueous altered materials on their surface  and in particular they show the 0.7 $\mu$m band in their spectra.

\section{Observations and Data Reduction}  

[HERE TABLE 1 ]

The data presented in this work were obtained  during 2 runs 
on March and November 1999 at the 1.52m telescope of the European Southern Observatory (ESO), in Chile.
The telescope was equipped with a Boller \& Chivens spectrograph and a Loral Lesser CCD as detector (2048$\times$2048 pixels). The grating
used was a 225 gr/mm, with a dispersion of 331\AA/mm in the first
order, covering the  0.42 $< \lambda <$ 0.93 $\mu$m spectral range. The CCD has a 15 $\mu$m square pixels, giving a dispersion of
about 5 \AA/pixel in the wavelength direction.\\
Each spectrum was recorded through a slit oriented in the East--West
direction.
The slit was opened from 2 to 5 arcsec in order to reduce effects due to differential refraction and the possibility of losing signal due to guiding errors of the telescopes. \\ 
An order sorting filter cutting the signal below 4200 \AA\ was used to prevent overlapping of the second spectral order on the spectrum.   \\
During each night, we also recorded bias, flat--field, calibration lamp, and solar analog star spectra at different
intervals throughout the night. The stars were observed at airmasses similar to those of the objects.

\begin{figure*}
\centerline{\psfig{file=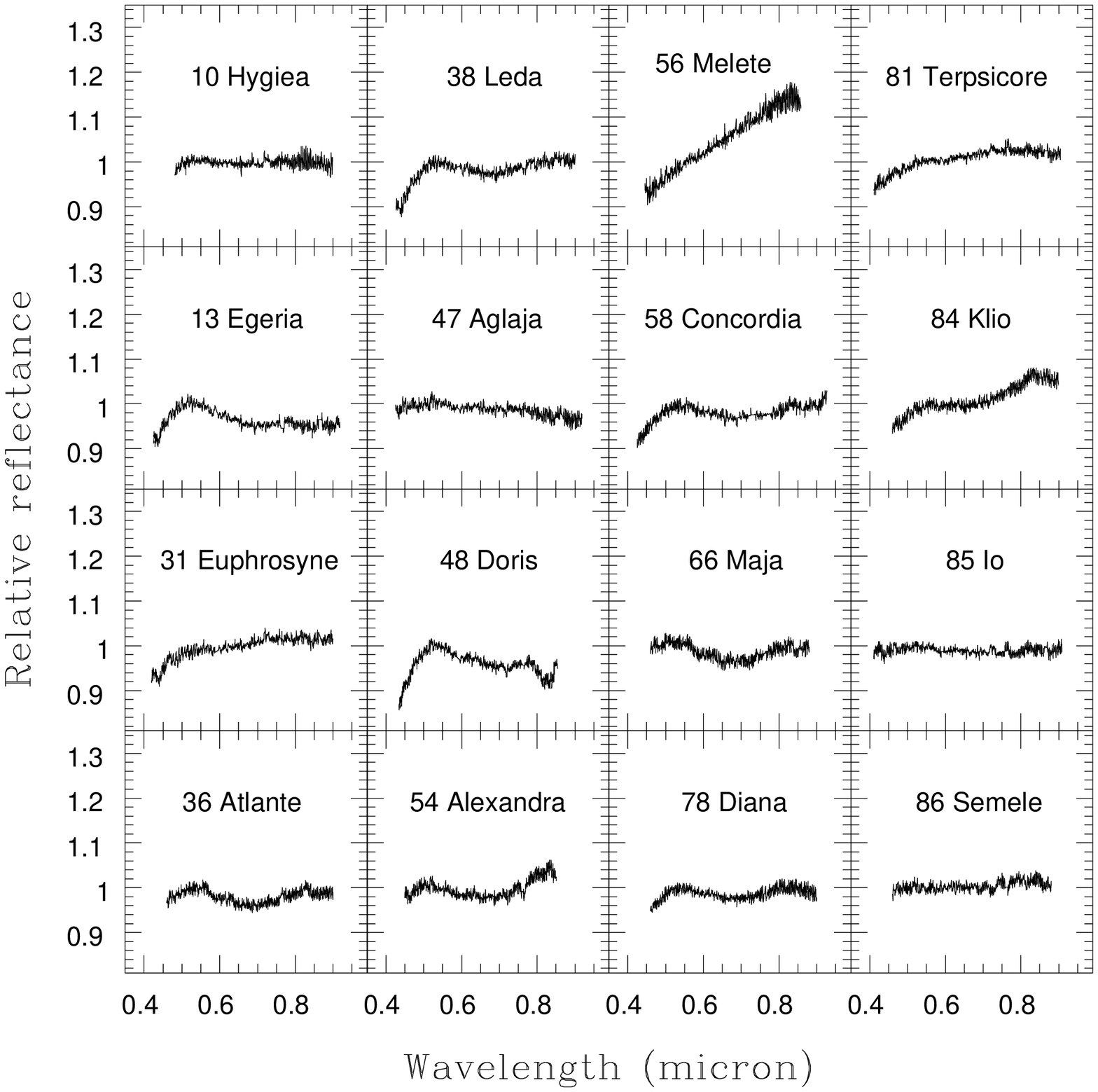,width=20truecm,angle=0}}
\caption{Reflectance spectra of the observed asteroids. The spectra are
normalized at 0.55$\mu$m.}
\label{ca}
\end{figure*}

\begin{figure*}
\centerline{\psfig{file=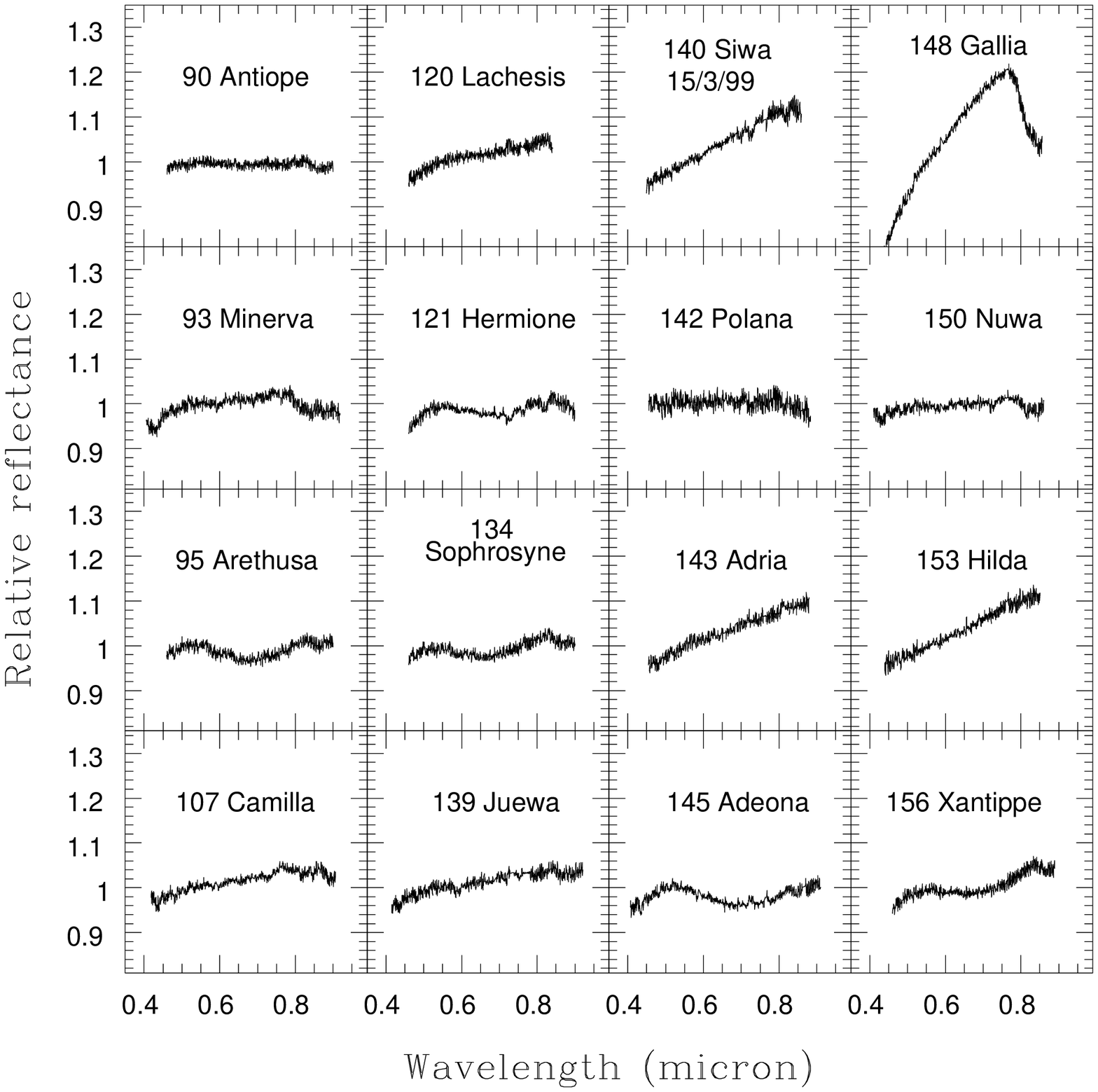,width=20truecm,angle=0}}
\caption{Reflectance spectra of the observed asteroids. The spectra are
normalized at 0.55$\mu$m.}
\label{cb}
\end{figure*}

\begin{figure*}
\centerline{\psfig{file=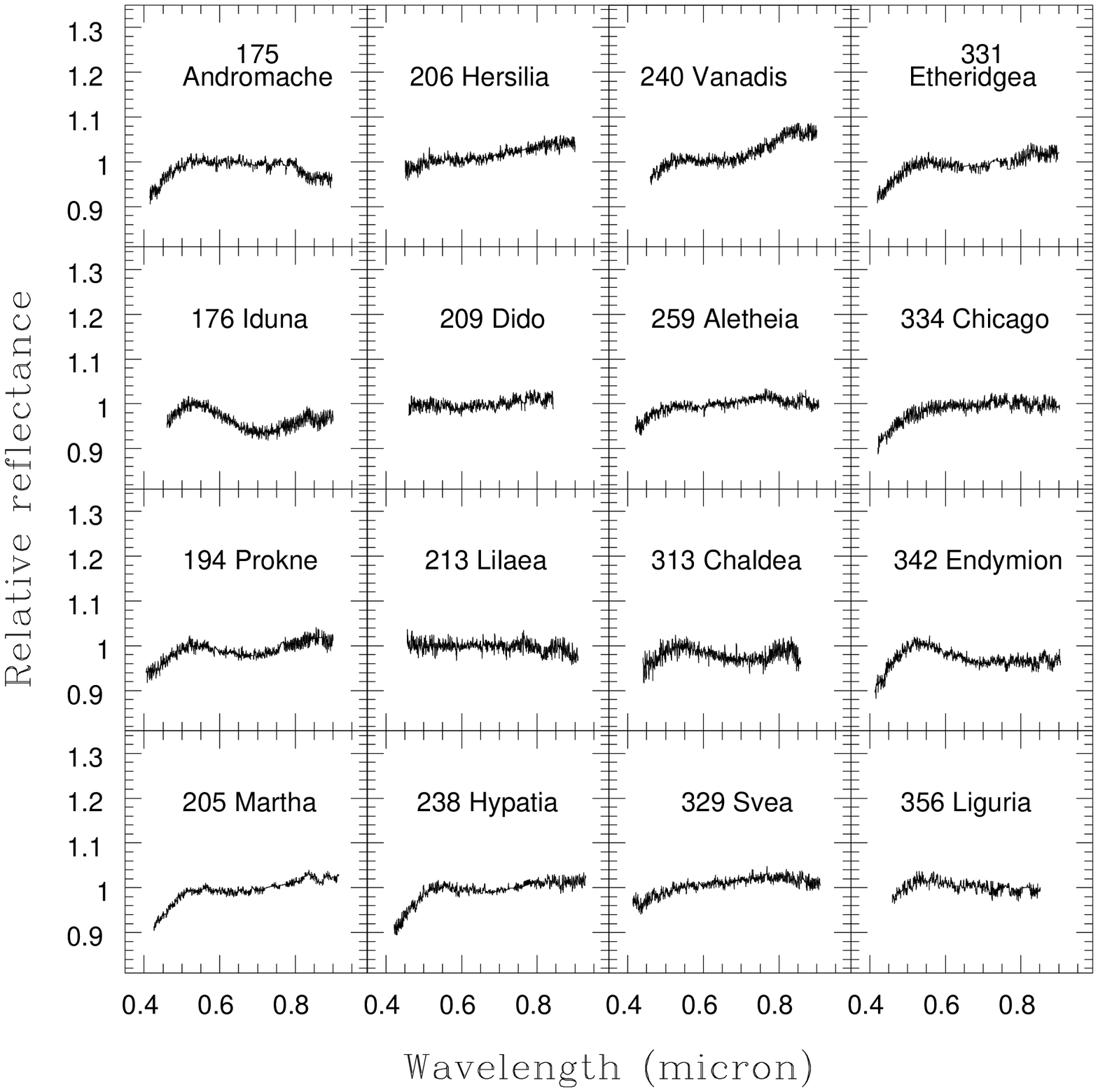,width=20truecm,angle=0}}
\caption{Reflectance spectra of the observed asteroids. The spectra are
normalized at 0.55$\mu$m.}
\label{cc}
\end{figure*}

\begin{figure*}
\centerline{\psfig{file=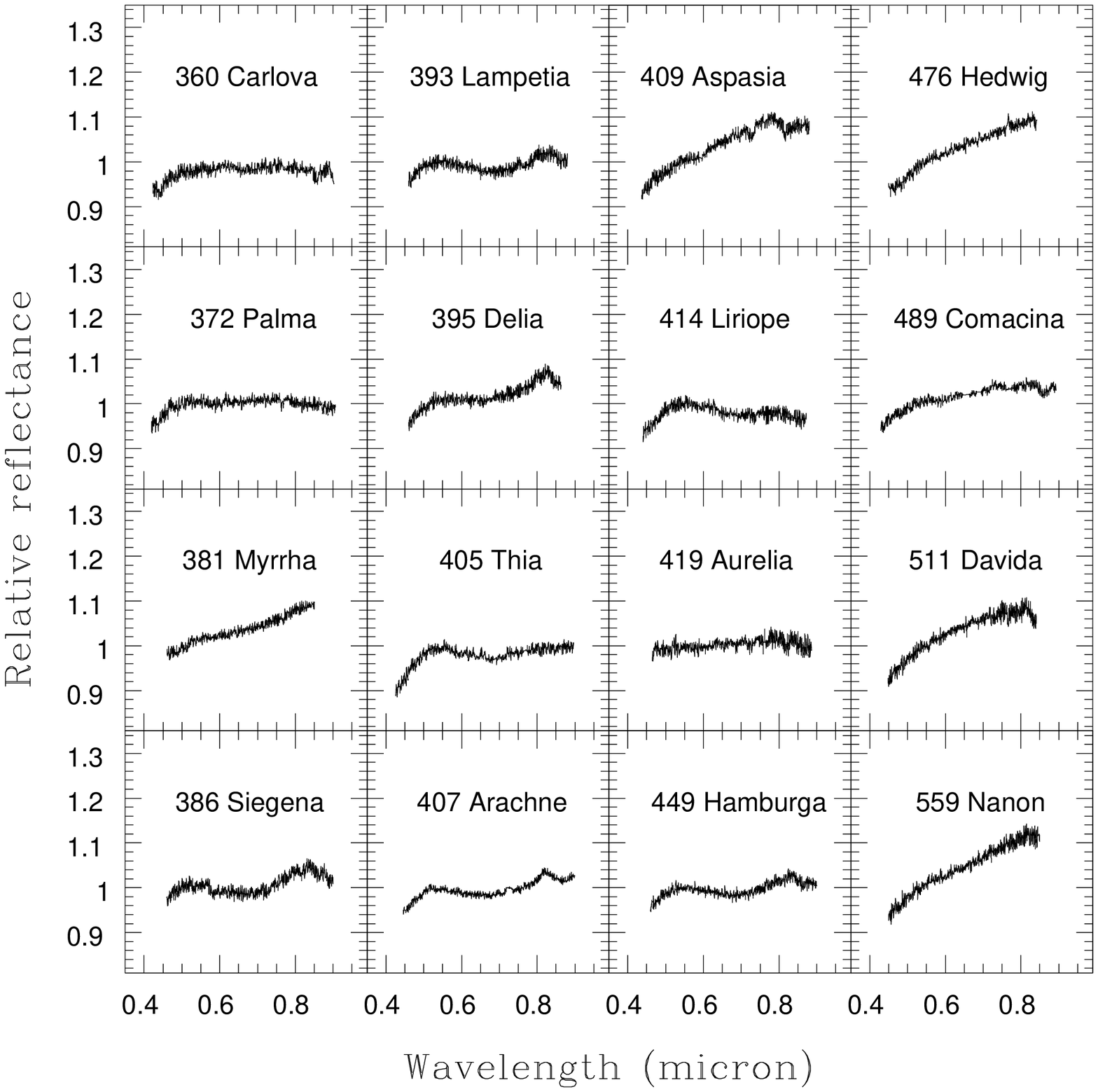,width=20truecm,angle=0}}
\caption{Reflectance spectra of the observed asteroids. The spectra are
normalized at 0.55$\mu$m.}
\label{cd}
\end{figure*}

\begin{figure*}
\centerline{\psfig{file=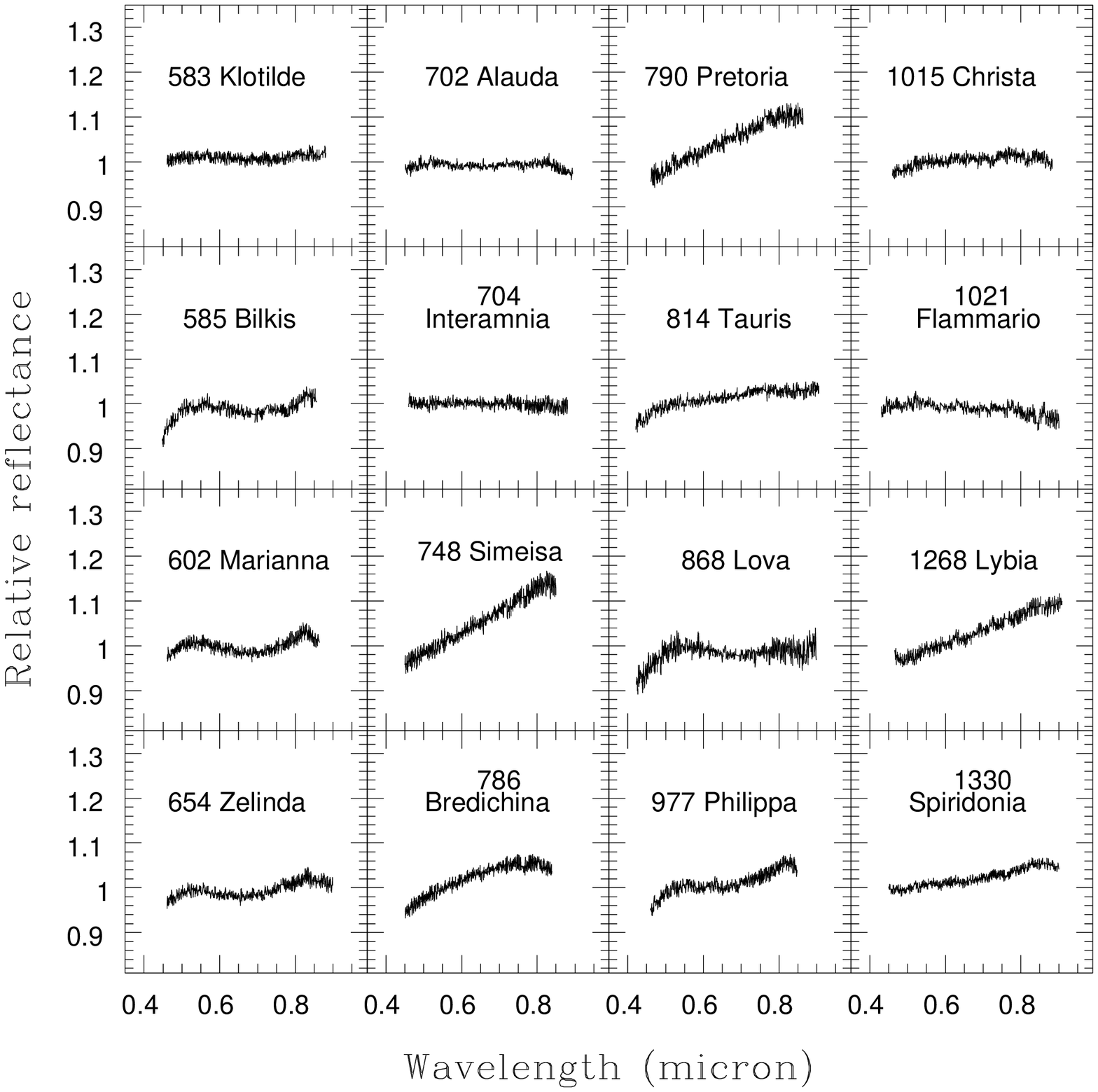,width=20truecm,angle=0}}
\caption{Reflectance spectra of the observed asteroids. The spectra are
normalized at 0.55$\mu$m.}
\label{ce}
\end{figure*}

The spectra were reduced using the software packages Midas as described by Fornasier et al. (1999, 2004b). The procedure includes the subtraction of the bias 
from the raw data, flat--field correction, cosmic ray removal, sky subtraction,
collapsing the two--dimensional spectra to one dimension, wavelength calibration,
and atmospheric extinction correction. 
Wavelength calibration was made using a lamp with He, Ar, Fe and Ne emission
lines. The residuals of the wavelength calibration were $\leq 3$ \AA.\\
The reflectivity of each asteroid was then 
obtained by dividing its spectrum by that of the solar analog star closest in
time and airmass to
the object. Spectra were finally smoothed with a median filter 
technique, using a box of 19 pixels in the spectral direction for each point of
the 
spectrum. The threshold was set to 0.1, meaning that the original value was
replaced by 
the median value if the median value differs by more than 10\% from the original
one.\\
Some spurious features due to incomplete removal of sky lines (in particular of
the O$_{2}$A band around 7600 \AA\ and of the water telluric bands around
7200 \AA\ and 8300 \AA)
are present on the asteroidal spectra. Anyway these
features are easily recognizable, and were disregarded in the spectral
analysis.\\
The spectra of the observed asteroids, all normalized at 0.55 $\mu$m, are shown
in Figures ~\ref{ca}--\ref{ce}. Details about the observational circumstances and the solar analogue stars used in the reduction process are reported in Table~\ref{tab1}.

[Here Figs 1, 2, 3, 4, 5]

[Here Table 1]

\section{Results}

For each asteroid we have looked for the presence of absorption features  associated with hydrated silicates and, when present, we have characterized the band center, depth, and width.
First, a linear continuum was fitted at the edges of the band, 
that is at the points on the spectrum outside the absorption 
band being characterized. Then the asteroid spectrum was divided 
by the linear continuum and the region of the band was fitted with a 
polynomial of order 2--4. 
The band center was then calculated 
as the position where the minimum of the spectral reflectance 
curve occurs (on the polynomial fit), and the band depth as the minimum of the polynomial fit (see Table~\ref{tab2}), following the method described in Fornasier et al. (1999). Errors in the band center and depth were evaluated from the standard deviation of several attempts in determining the continuum and using different polynomials in the fit. \\ 
The slope of the asteroid spectrum between 0.55-0.80 $\mu$m was also calculated 
 and reported in Table~\ref{tab2}. The slope error bars take into account the
$1\sigma$ uncertainty of the linear fit plus 0.5\%/$10^3$\AA\ attributable to the spectral variation due to the use of different solar analog 
stars during the night.

[HERE TABLE 2]

\subsection{ {\bf Absorption features in the visible range}}

In Table~\ref{tab2}  we report the results of this analysis for the asteroids presented in this paper (80), together with those observed during the same observational campaigns and presented in Fornasier et al. (1999), for a total of 110 primitive asteroids. Within this sample we have 87 C-type, 6 G-type, 6 F-type, and 10 P-type asteroids. Asteroid 148 Gallia, originally classified as GU, is clearly an S-type asteroid (see Fig.~\ref{cb}) so it was excluded from our analysis.\\
We considered only the absorption features deeper than the peak-to-peak 
scatter (that is depth $>$ 0.8\%) in the spectrum, which, from 
previous experience, seems to 
be a better indicator of the spectrum quality than the calculated 
signal to noise ratio (Vilas \& Smith, 1985). 
The depth of aqueous alteration  bands within the sample of 109 asteroids varies between 1\% and 7\% with respect to the continuum. The principal band identified is the one centered around 0.7 $\mu$m, that is  associated with $Fe^{2+}\rightarrow Fe^{3+}$ charge transfer absorptions in phyllosilicate minerals (Vilas \& Gaffey, 1989; Vilas et al., 1993). This band is often associated with an evident UV absorption below $\sim$ 0.5 $\mu$m, due to a strong ferric oxide intervalence charge transfer transition (Vilas et al., 1994). \\
Other features identified on some spectra are: the 0.43 $\mu$m band due to the $^{6}A_{1}\rightarrow ^{4}A_{1}, ^{4}E(G) $ spin-forbidden crystal field transition assigned to ferric iron (Hunt \& Ashley, 1979; Townsend, 1987) which Vilas et al. (1993) attributed to jarosite, a secondary product of the aqueous alteration of iron sulfide minerals such as pyrite (Burns, 1987); the 0.8-0.9 $\mu$m band due to the $^{6}A_{1}\rightarrow ^{4}T_{1}(G) ~\ Fe^{3+}$ charge transfer transition in iron oxides (Hunt \& Ashley, 1979; Townsend, 1987); the 0.6 $\mu$m band attributed to $^{6}A_{1}\rightarrow ^{4}T_{2}(G) ~\ Fe^{3+}$ charge transfer transition in iron oxides (Feierberg et al., 1985; Vilas et al., 1994).  \\
We find that about 65 \% of the C-type  and all the G-type asteroids investigated in this survey reveal features suggesting the presence of hydrous materials, while we do not see  evidence of
hydrated materials on the spectra of the other low albedo asteroids (B, F, and P-types) investigated.

\subsection{ {\bf Correlations between the 0.7 and 3 $\mu$m bands}}

[HERE TABLE 3]

We compared the visible spectra of the observed asteroids with the 3 $\mu$m data available in the literature. In fact it is known that the 0.7 $\mu$m band and the 3 $\mu$m band, attributed to OH and overlapping overtones of H$_2$O, are related (Vilas et al., 1993; Vilas, 1994).
We found that 37 of our investigated asteroids have also been observed in the IR region by other authors (Table~\ref{tab3micron}). Comparing this sample with our visible data, we found a strong correlation between the visible 0.7 $\mu$m band due to hydrated materials and the 3 $\mu$m one: 95\% of the asteroids (19 out of 20 objects) having the 0.7 $\mu$m band show also the 3 $\mu$m band (all except 444 Gyptis).  Five objects do not show hydrated mineral features at all, neither in the visible nor in the  NIR range. On the other hand, the 3 $\mu$m band is not always  associated with the 0.7 $\mu$m one (Table~\ref{tab3micron}): 6 asteroids have silicate hydrated features in the IR range but not in the visible range, and 6 have a round shaped 3 $\mu$m band  associated with H$_2$O frost rather than to hydrated silicates (Takir \& Emery, 2012), so they normally did not experience aqueous alteration in the past.\\
If the 0.7 $\mu$m band is present on the spectra, it is almost always accompanied  by the 3 $\mu$m feature. Nevertheless, if the 0.7 $\mu$m band is not seen, the 3 $\mu$m band may still be present on the spectra. It must be noted that the 0.7 $\mu$m band is much fainter than the 3 $\mu$m one, so it may be easily hidden in low S/N ratio spectra. \\
The aqueous alteration process on iron bearing silicates results in hydrated minerals having both the 0.7 and 3 $\mu$m absorption bands. Alternatively, asteroids having the 3 $\mu$m band but not the 0.7 $\mu$m one may be iron poor, or may have converted all the Fe$^{2+}$ in Fe$^{3+}$, or may have experienced mild
heating episode (400$^{o}C <$ T $<600^{o}C$), occurring after the aqueous alteration, which decreased the 0.7 $\mu$m band depth. Indeed some heating experiments on Murchinson (CM2) samples (Hiroi et al., 1996; Cloutis et al., 2011a) show  that the 0.7 and 3 $\mu$m bands are both present up to 400$^{o}$C, and both disappear for T $>$ 600$^{o}$C as minerals become completely dehydrated. Between 400$^{o}$C and 600$^{o}$C, the 0.7 $\mu$m band weakens and disappears, and the 3 $\mu$m band gets shallower. 

Our results confirm those found by Vilas (1994) and Takir \& Emery  (2012) who saw both the 0.7 and 3 $\mu$m features on $\sim$ 60-80\% of the investigated bodies on a sample of 27-29 asteroids observed both in the visible and near infrared wavelength. Also Howell et al. (2011) confirmed the strong correlation between the presence of the 0.7 and 3 $\mu$m band on a larger sample of 156 asteroids belonging to the C- and X-complex.  \\
This analysis indicates that the study of the 0.7 $\mu$m band provides only a lower limit on the number of C-complex hydrated asteroids in the main belt.


\section{Aqueous alteration: the big picture}

[HERE TABLE 4]

[Link to TABLE 5 that must appear only in electronic form]

To fully investigate the aqueous alteration process, we enlarge our sample of 109 asteroids including the visible spectra of primitive asteroids available in the literature. A total sample of 600 visible spectra belonging to the primitive classes C, G, B, F, and P has been collected from several published surveys: SMASS I and II (Xu et al., 1995; Bus \& Binzel, 2002a), and S30S2 (Lazzaro et al., 2004) surveys, and from dedicated observing campaigns on aqueous alteration (Fornasier et al., 1999; Sawyer et al., 1991; Vilas \& McFadden, 1992; Vilas et al., 1993; this work). We report in Table~\ref{summary} a summary on all the investigated asteroids for the different taxonomic classes with the related percentage of hydrated objects and albedo values.\\ 
All these spectra have been re-analyzed adopting the same procedure presented before, and we have characterized the 0.7 $\mu$m band, when present: band center position, width and depth compared to the continuum. \\
When multiple observations of an asteroid are available in the literature, we considered it hydrated if we detect aqueous alteration bands in at least one of the available spectra (with S/N ratio good enough). When all the literature spectra of a given asteroid show aqueous alteration bands, we check spectral consistency and consider the results from the data with the highest S/N ratio. \\
The complete list of the asteroids analyzed in the literature and the related physical parameters and band absorption characterization is presented in Table~\ref{tab_all}. The errors in band center, depth and slope presented in Table~\ref{tab_all} were simply computed from the fitting process without additional contribution derived from the spectrum quality check. Indeed, for the data from the literature we do not know the quality of the observing nights and the consistency of the solar analogs spectra used to obtain the asteroid reflectance. Moreover the spectra come from different databases with non homogeneous S/N ratio quality. 

We ran a Spearman Rank Correlation (Spearman, 1904) to search for possible
correlations between the hydrated asteroids (from Tables~\ref{tab2} and~\ref{tab_all}) and the object's albedo, diameter, taxonomic class and orbital elements. The Spearman correlation function gives a two-element vector containing the rank correlation coefficient ($\rho$) and the 
two-sided significance of its deviation from zero ($P_{r}$).
The value of $\rho$ varies between -1 and 1; if $|\rho|$ is close to zero, then
there is no correlation and if $|\rho|$ is close to 1, then a correlation
exists. The significance ($P_{r}$) is a value in the interval $0 < P_{r} < 1$. A
small value indicates a significant correlation.
We consider a strong correlation to have $P_{r} < 0.01 $ and $|\rho| > 0.6 $,
and a weak correlation to have $0.01 < P_{r} < 0.08 $ and $ 0.3 < |\rho| < 0.6 $. \\

We want to emphasize that the following analysis of aqueous alteration on primitive main belt asteroids is made only considering the visible region. As we explain in the previous section, the stronger hydratation band at 3 $\mu$m may be present even if the 0.7 $\mu$m is not seen. This means that the results of our analysis give only a lower limit in the hydration state of primitive main belt asteroids. 
 
\subsection{{\bf Spectral variability}}

We compare  the results on the 109 asteroids observed in our survey (Table~\ref{tab2}) with those presented in the literature (Table~\ref{tab_all}). Generally there is a good agreement, that is the 0.7 $\mu$m band is seen by us and other authors or the spectrum of a given asteroid is featureless in all the available spectra.
Nevertheless, there are some cases where the 0.7 $\mu$m band is seen in the spectrum of a given asteroids but not in other datasets: 10 Hygiea, 24 Themis, 41 Daphne, 54 Alexandra, 65 Cybele, 84 Klio, 90 Antiope, 194 Prokne, 240 Vanadis, 356 Liguria, 386 Siegena, 393 Lampetia, 414 Liriope, 444 Gyptis, 702 Alauda, 712 Boliviana, and 776 Berbericia. Considering that we are comparing different datasets, and that some spectra from the literature are sometimes noisy, it is possible that some spectral differences are related to observational uncertainties and/or differences in data acquisition and reduction. It must be noted that most of the aforementioned asteroids are large bodies (most are bigger than 120 km in diameter), and it is possible that they really display surface variability. These objects are good candidates for further spectroscopic observations at high S/N ratio covering the full rotational period to investigate and eventually confirm  their surface heterogeneity.

\subsection{{\bf Aqueous alteration and the taxonomic classes}}

\begin{figure}[h]
\centerline{\psfig{file=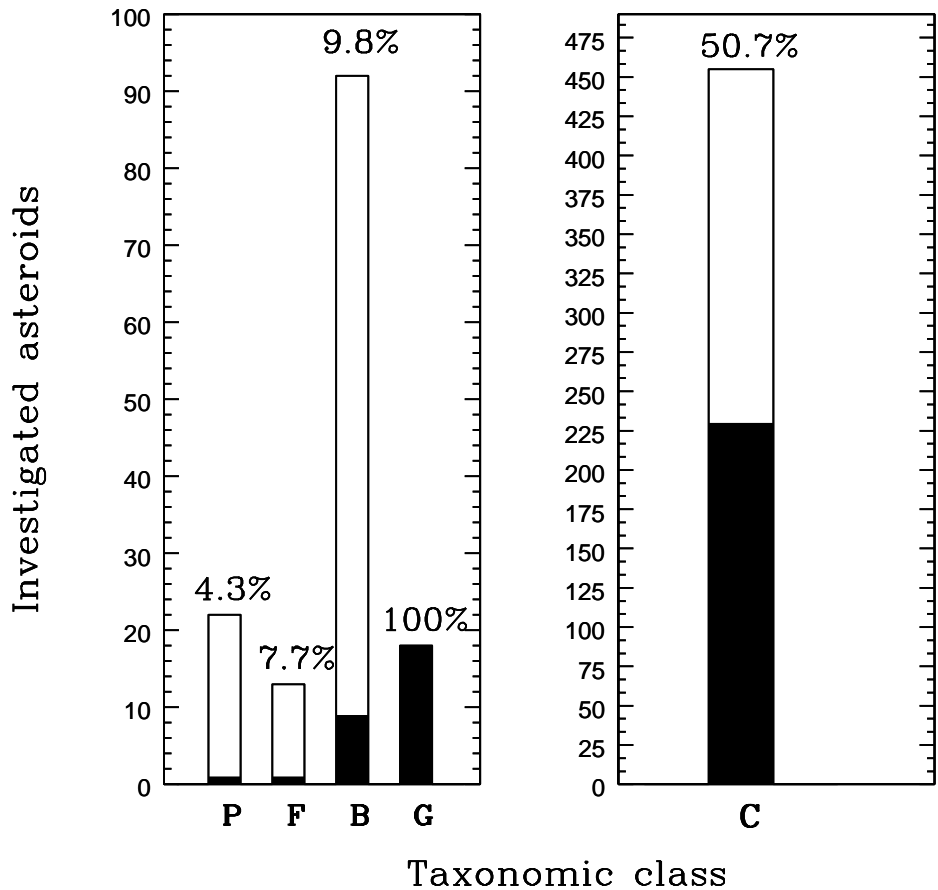,width=14truecm,angle=0}}
\caption{Number of investigated asteroids as a function of their taxonomic class.
The black part represents the hydrated asteroids as respect to the
whole sample for each taxonomic class. The percentage of hydrated asteroids
for each class is also reported.}
\label{classi}
\end{figure}

Using an algorithm to detect the 0.7 $\mu$m band from ECAS photometry, Vilas (1994) found a relative incidence of this feature of 85.7\% in G, 47.7\% in C, 33\% in B, 16.5\% in F and 8.3\% 
in P-type asteroids. Later works based on spectroscopic data revealed a more extensive incidence of aqueous alteration: 60-65\% 
of the C-type investigated (Barucci et al. (1998), and Fornasier et al. (1999) on a sample of 29 and 34 objects, respectively), a percentage in good agreement with the fraction of C-types that present the 3 $\mu$m band (Rivkin et al., 2002).
In our sample of 600 primitive asteroids observed in the visible range (Fig.~\ref{classi}), we find 230 out of 455 C-types having absorption bands  associated with the hydrated silicates, 1 (65 Cybele) out of 22 P-type, 1 out of 13 F-type, 9 out of 92 B-type, and the totality of the 18 G-type asteroids (Table~\ref{summary}). So the percentage of hydrated asteroids from this investigation in the visible range is 4.3\% for the P, 7.7\% for the F, 9.8\% for the B, 50.7\% for the  C, and 100\% for the G-type asteroids. 

[HERE FIG. 6]

These data confirm the presence of an aqueous alteration sequence as already suggested by Vilas (1994) that starts from the P-type, primitive and unaltered, and increases through the P $\rightarrow$ F $\rightarrow$ B $\rightarrow$ C $\rightarrow$ G asteroids, these last being heated to the point where melted ice could cause pervasive aqueous alteration. \\
Analyzing the photometric data of the SDSS survey, Rivkin (2012) found a lower percentage (30$\pm$5\%) of hydrated C-complex asteroids showing an absorption feature in the 0.7 $\mu$m region. It must be noted that the comparison between our and his results is not simple because in his analysis Rivkin included in the C-complex different taxonomic types following the Bus \& Binzel (2002a) taxonomy (B, C, Cb, Cg, and the hydrated Cgh and Ch-types), while we used the Tholen taxonomy, and he used 5 broadband filters from the SDSS survey instead of spectroscopic data. Moreover, Vilas (2005) tried to use SDSS data to look for the presence of 0.7 $\mu$m band on asteroids surfaces, and concluded that the SDSS filters band center and width are not optimized for the detection of this faint and shallow absorption feature.  This may explain why the percentage of hydrated asteroids according to Rivkin (2012) is lower than the one we obtain. If we consider the whole sample of the Tholen B, F, G and C-type asteroids we investigated in the visible region, classes which belong to the C-complex according to the Bus \& Binzel (2002a) taxonomy, then we find a percentage of 45\% of hydrated C-complex asteroids. According to the correlation between the 0.7 and 3 $\mu$m bands (Howell et al. 2011),  the number of C-complex asteroids having the 3 $\mu$m band and so hydrated silicates on their surface must be around 1.5 times bigger, that is of the order of 70\%.

\subsection{{\bf Location of the aqueous alteration region}}

[HERE FIGURE 7 and 8]

\begin{figure}[h]
\centerline{\psfig{file=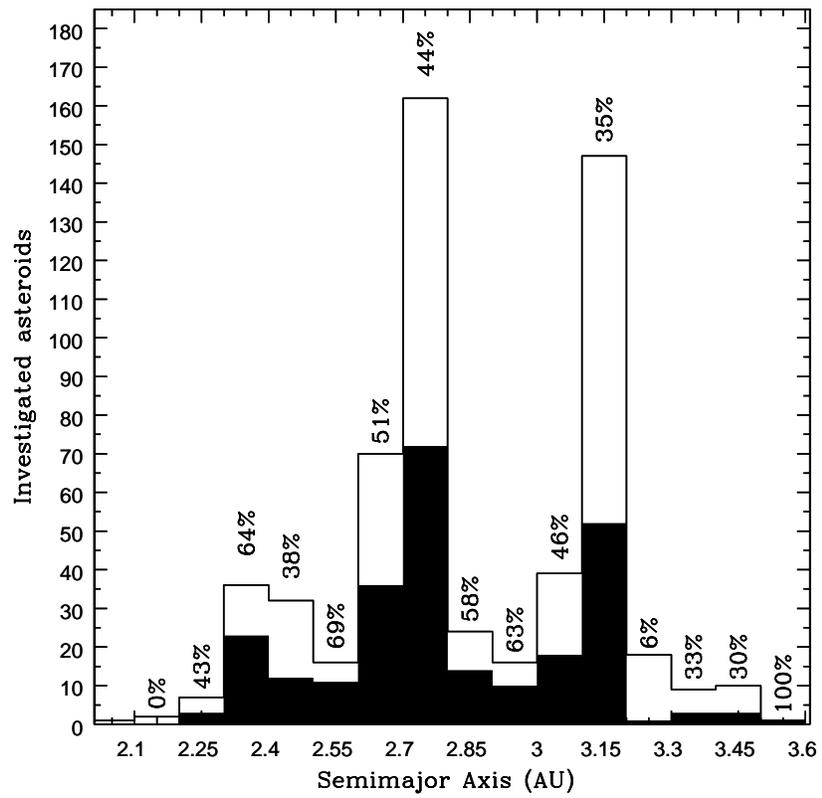,width=12truecm,angle=0}}
\caption{Number of the investigated objects as a function of the semimajor axis.
The black part represents the hydrated asteroids.}
\label{isto}
\end{figure}

Vilas (1994) defined the aqueous alteration zone, that is the region where the process dominates in primitive asteroids,  between 2.6 and 3.5 AU. 
McSween et al. (2002) and Grimm \& McSween (1993) presented a quantitative model to explain the radial thermal structure of the asteroid belt. In their model they identify the $\sim$ 2.7 AU region as the approximate
distance for the transition from melted or metamorphosed asteroids to those that experienced aqueous alteration. According to their models, melted ices must have been produced within 2.7 and 3.4 AU, with asteroids located further than 3.4 AU being unaltered because ice was never melted.\\
In our sample of primitive asteroids  we observe a dominance of hydrated bodies between 2.3 and 3.1 AU (in semimajor axis). This result suggests that the zone of aqueous alteration extends to the inner belt, and that the aqueous alteration process is less effective beyond 3.1 AU (Fig.~\ref{isto}). The fraction of hydrated asteroids decreases with increasing semimajor axis from the middle of the asteroid belt going outward. A similar distribution has been also found by Carvano et al. (2003), who analyzed the visible spectra of primitive asteroids from the S3OS2 survey, and by Takir \& Emery (2012), who investigated the 3 $\mu$m band on primitive asteroids. The decrease of the abundance of hydrated C-type asteroids at large heliocentric distance is also confirmed by Rivkin (2012) from the SDSS survey, who found that hydrated C asteroids are preferentially concentrated in the mid asteroid belt, reaching a minimum in the outer belt. \\
If we consider the distribution of C-type asteroids versus their perihelion distance ($q$), we find that the aqueous alteration process dominates (the percentage of hydrated asteroids is $>$ 50\%) for 1.8 $< q <$ 2.6 AU, and it is less effective for higher q, except at 3.1 $< q <$ 3.3 AU, where the percentage rises up again but the population has few objects. The Spearman test gives a faint anticorrelation between the percentage of hydrated C-type asteroids and the perihelion distance, with a correlation coefficient $\rho$ = -0.4755 and a significance level $P_r$= 0.0163 (if we consider the whole sample of asteroids then the correlation is slightly weaker: $\rho$=-0.4157 and $P_r$ = 0.0388). \\
Takir \& Emery (2012) found a weak anticorrelation between the heliocentric distance and the 3 $\mu$m band depth, suggesting that hydrated silicates possibly decline in abundance with the heliocentric distance. Analyzing our sample, we find no correlation between the 0.7 $\mu$m band depth and the heliocentric distance neither in the whole sample nor in the individual taxonomic classes.

In this research we limited our analysis to the main belt asteroids. For the Near Earth population, possible bands  associated with hydrated materials have been found on very few objects up to date. Binzel et al. (2004) found the 0.7 $\mu$m band on 2099 Opik, indeed classified as a Ch-type in the Bus taxonomy. The NEO (175706) 1996 FG3 has an absorption feature in the 3 $\mu$m range diagnostic for hydrated/hydroxylated minerals on its surface (Rivkin et al., 2013), but no bands  associated with hydrated materials were found in the visible region (Perna et al., 2013; De Leon et al., 2013). The NEO 1999 JU3, target of the Hayabusa-2 mission, has an absorption band centered around 0.7 $\mu$m attributed to hydrated silicates by Vilas (2008); nevertheless, this band was seen in only one out of the 3 spectra acquired during different observing nights by Vilas (2008), and it was not found in any of the spectra obtained by Lazzaro et al. (2013), spectra which cover 70\% of its surface. The fact that hydrated primitive asteroids are rare among the NEO population indicates that probably their surfaces underwent heating episodes in the past which removed the 0.7 $\mu$m band, or that their source is not representative of the whole main belt C-complex. 

For asteroids located at larger heliocentric distances, no bands clearly associated with the aqueous alteration process have yet been detected on Jupiter Trojans, of either large or small sizes (Jewitt \& Luu, 1990; Lazzaro et al., 2004; Fornasier et al., 2004b; Bendjoya et al., 2004; Dotto et al., 2006; Emery et al., 2006; Fornasier et al., 2007; Melita et al., 2008; de Luise et al., 2010; Emery et al., 2011), even if water ice is considered to be abundant in their interior. Fornasier et al. (2007) detected a peculiar family, the Eurybates family in the L4 swarm, abundant in spectrally flat objects, similar to Chiron-like Centaurs and C-type main belt asteroids. Some of the Eurybates members showed a  drop--off of the reflectance shortward of 5200 \AA\, attributed to the intervalence charge transfer transition in oxidized iron, that may be associated with, but 
not uniquely indicative of, phillosilicates. Indeed, no other absorption bands attributed to hydrated silicates   (or water ice) were reported for these targets  in the visible and near infrared ranges (Fornasier et al., 2007; de Luise et al., 2010).

At further distances, few Centaurs and TNOs (2000 GN171, Huya, Chariklo, 2003 AZ84) show faint absorption features in the visible range that were attributed to the aqueous alteration process (Lazzarin et al., 2003; De Bergh et al., 2003; Fornasier et al., 2004a, 2009; Alvarez et al., 2008). At these large heliocentric distances the aqueous alteration process may be driven by the hydrocryogenic alteration which may have altered mixtures of dust and ice in thin interfacial water layers (Rietmeijer \& MacKinnon, 1985), or by radiogenic heating and impacts that may have produced enough heating to make this process effective.  Finally, considering that hydrated silicates have been detected in IDPs (Mackinnon \& Rietmeijer 1987), it is also possible that aqueous alteration products could have existed on grains when the nebula started to cool down.

\subsection{{\bf Aqueous alteration and asteroids diameter}}

\begin{figure}[h]
\centerline{\psfig{file=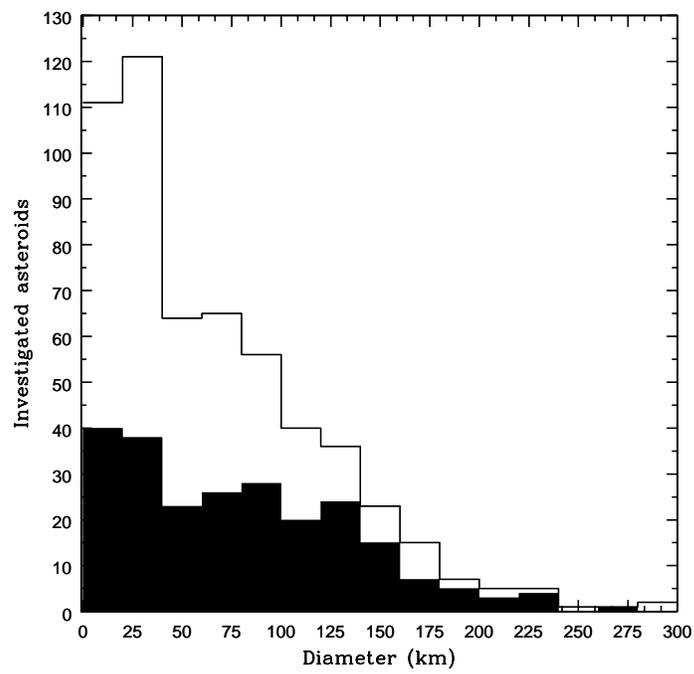,width=10truecm,angle=0}}
\caption{Number of the investigated asteroids as a function of the asteroids diameters.
The black part represents the hydrated asteroids.}
\label{istodiam}
\end{figure}

\begin{figure}[h]
\centerline{\psfig{file=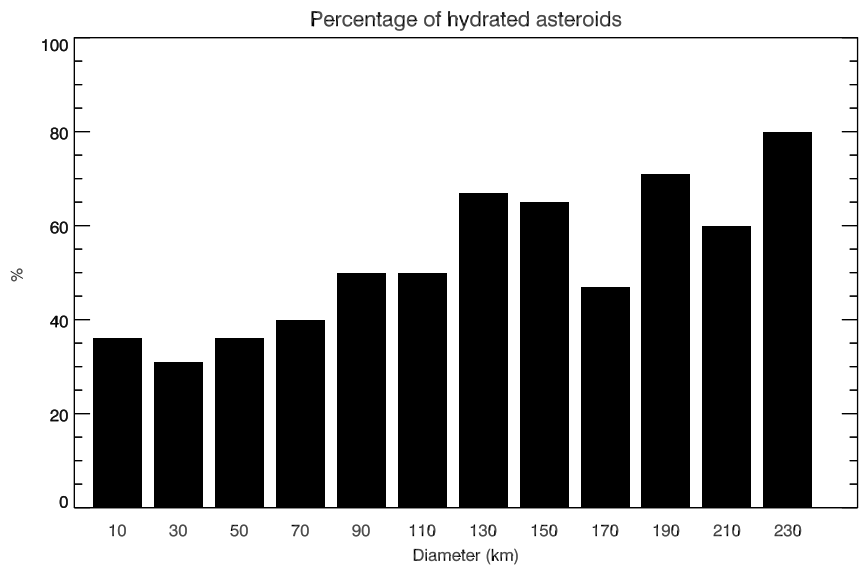,width=13truecm,angle=-90}}
\caption{Percentage of the hydrated asteroids as a function of their diameters.}
\label{percdiam}
\end{figure}

Most of the investigated asteroids (557 out of the 600) have albedo and diameter values mainly from the WISE or the IRAS surveys (Tables~\ref{tab2}, and~\ref{tab_all}). We studied the percentage of hydrated asteroids versus the diameter, considering a bin size of 20 km and excluding the few largest asteroids having D$>$ 240 km, that are all hydrated. We find that the percentage of hydrated asteroids and the size are strongly correlated (Figs.~\ref{istodiam}, and~\ref{percdiam}). The Spearman rank test gives a strong correlation coefficient $\rho$ = 0.8626 for the percentage of hydrated asteroids versus their diameter, with a significance level $P_r$=0.0004. Considering a smaller bin size of 10 km, these two quantities are still correlated but with a smaller coefficient value ($\rho = 0.6920$ and $P_r = 0.0002$). \\
This analysis indicates that the percentage of hydrated asteroids decreases with the size and that the aqueous alteration process dominates for primitive asteroids with 50 $< D <$ 240 km, as claimed previously in the literature (Howell et al., 2001; Jones et al., 1990; Vilas \& Sykes, 1996). In particular, the asteroids larger than 100 km are strongly affected by the aqueous alteration process, as more than half of them present absorption bands  associated with hydrated silicates. These large asteroids must have retained water ice in their interior and must have produced internal heating sufficient enough to melt the ice. Liquid water would have then reached their surfaces by hydrothermal circulation, and reacted with the surface to produce hydrated minerals. \\
Within the heliocentric range of the main asteroid belt where most C-class asteroids are found, asteroids with diameters above 20 km should have been
heated to the point of water mobilization (Herbert et al., 1991; Grimm \& McSween, 1993; Shimazu \& Teresawa, 1995). According to McSween et al. (2002), icy bodies with radii $<$50 km would have experienced aqueous alteration and metamorphism within about 1 m.y. of accretion. Aqueous alteration on much larger bodies would have been delayed by about 5 m.y. or more relative to the time of accretion.

As explained by Vilas \& Sykes (1996), according to the models of proposed heating mechanisms, asteroids approaching 50-km diameter
have been heated to temperatures exceeding the laboratory temperatures at which thermal metamorphism began reducing the depth of the 0.7 $\mu$m feature, as seen in heated Murchinson carbonaceous chondrite samples, while hydrated silicates are supposed to survive on primordial large asteroids. 
Thus smaller C-complex asteroids we observe nowadays are expected to be the fragments of hydrated larger asteroids, while the largest asteroids could be battered remnants of asteroids that originally underwent aqueous alteration. The fact that the percentage of hydrated asteroids decreases with the size is consistent with the hypothesis that the primordial parent bodies of the smaller asteroids have been collisionally disrupted and scattered, and the parent bodies of the larger asteroids have been collisionally disrupted but still bound together by gravitational attraction.

[HERE FIG. 8 and 9]

\subsection{{\bf Aqueous alteration and asteroids albedo}}
[HERE FIG 10 and 11]

We report in Table~\ref{summary} a summary of the number of observed asteroids, percentage of hydrated bodies and related mean albedo values (with the associated standard deviation) for each taxonomic type investigated. G-type asteroids, that experienced extensive aqueous alteration, have indeed the highest mean albedo. Nevertheless, less than 10\% of the B-type asteroids show signatures of hydrated silicates, and their mean albedo value is higher than that of the C-types, half of which have the 0.7 $\mu$m absorption band. If we look at the B, C and G-types, where more than one asteroid show signatures associated with the hydrated silicates, we do not see any differences in the albedo values from hydrated and non hydrated asteroids within the same taxonomic class.

\begin{figure}[h]
\centerline{\psfig{file=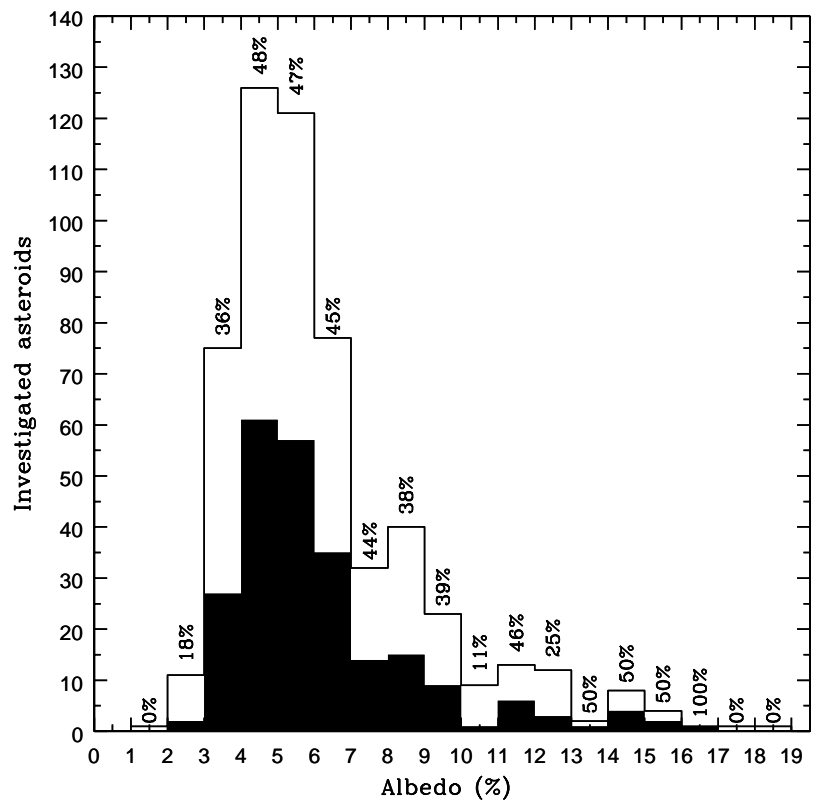,width=12truecm,angle=0}}
\caption{Number of the investigated asteroids as a function of the geometric albedo.
The black part represents the hydrated asteroids.}
\label{istoalbedo}
\end{figure}

\begin{figure}[h]
\centerline{\psfig{file=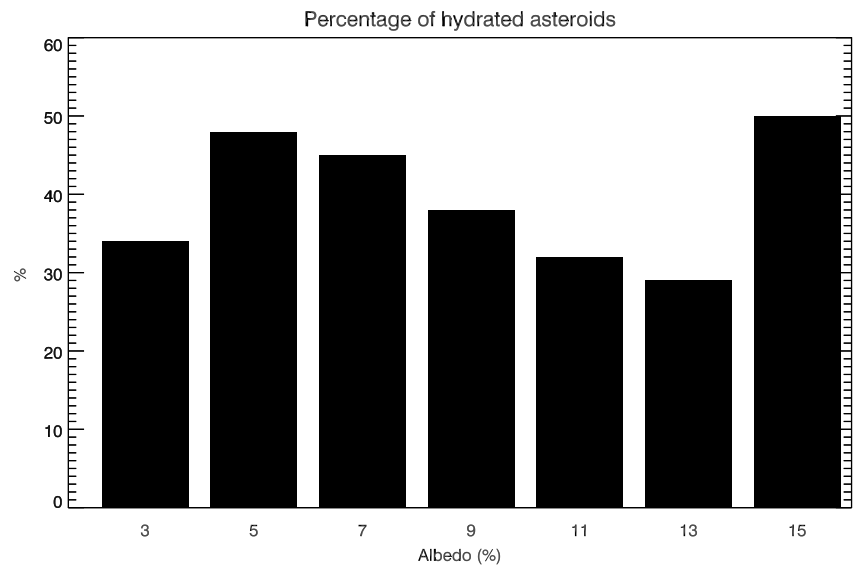,width=13truecm,angle=-90}}
\caption{Percentage of hydrated asteroids as a function of the geometric albedo.}
\label{percalbedo}
\end{figure}

Vilas (1994), analyzing ECAS data and using the mean albedo value for the P, B, C, and G classes derived from IRAS observations, identified a correlation between the asteroids' albedo and the aqueous alteration process, finding that the percentage of the observed hydrated asteroids grew for increasing albedo values. This correlation was explained with the progressive leaching of iron
from silicates as the aqueous alteration proceeds. Leached iron (iron is the
most important opaque phase in the visible range  associated with aqueous
alteration process) would be enveloped into magnetite and iron sulfide
grains, so less material would be available to absorb the incoming
sunlight and this would cause the increasing of the albedo (Vilas, 1994).
Using the individual albedo values available from WISE and IRAS, we do not find any correlation between the percentage of hydrated asteroids and their albedo, as shown in Figs.~\ref{istoalbedo} and~\ref{percalbedo}. Indeed the Spearman rank test, run on an albedo bin size of 0.02, gives no correlation between the albedo and the percentage of hydrated asteroids ($\rho= -0.0358$ and $P_r=0.9394$).\\
The percentage of hydrated asteroids increases from very dark surfaces to albedo values of 4-6\%, but then it decreases for higher albedo values, except for few relatively bright asteroids having 14$ < p_v < $16\%. We can conclude that no correlation is found between albedo and hydration state.

\subsection{\bf Comparison between CM chondrites and hydrated asteroids}

\begin{figure}[h]
\centerline{\psfig{file=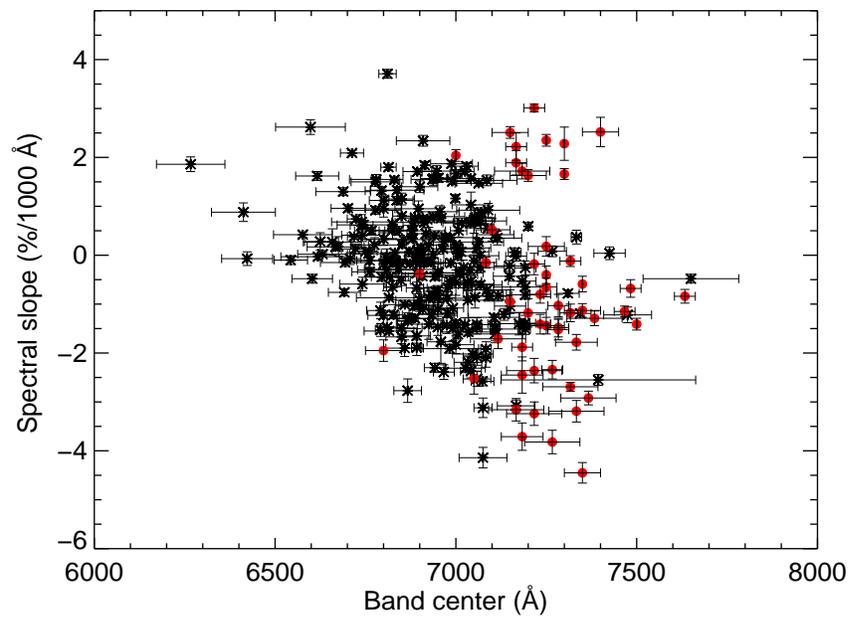,width=12truecm,angle=0}}
\caption{Visible spectral slope (in the 0.55-0.80 $\mu$m range) versus band center for the hydrated primitive asteroids (in black) and the CM meteorites investigated (red circles).}
\label{metcenslope}
\end{figure}

\begin{figure}[h]
\centerline{\psfig{file=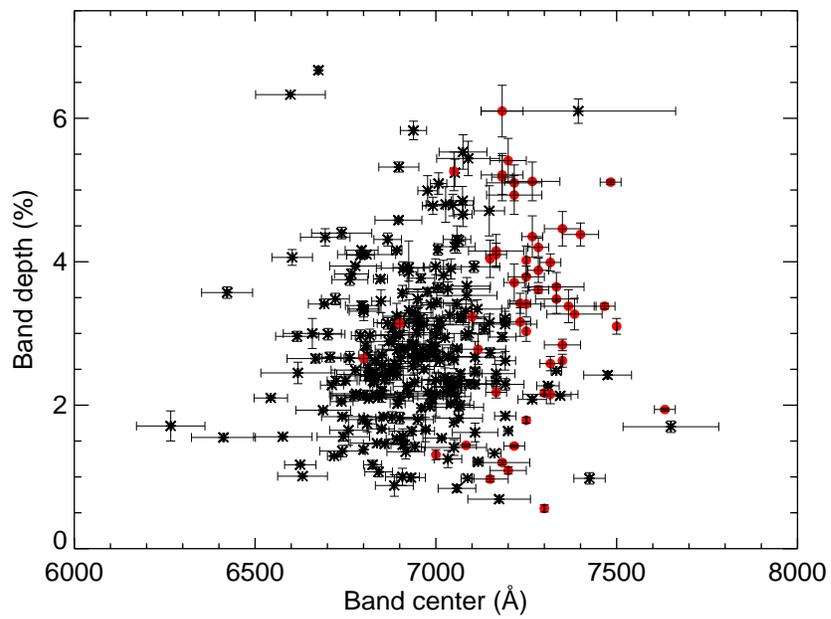,width=12truecm,angle=0}}
\caption{Band depth versus band center for the hydrated primitive asteroids (in black) and the CM meteorites investigated (red circles). }
\label{metcendepth}
\end{figure}

In the meteorite collection, hydrated minerals are found mostly among carbonaceous chondrites such as CI, CM and CR types, whose mineralogies indicate a low level of metamorphism ($<$ 1200 $^o$C) and evidence for aqueous alteration. 
Studying the phyllosilicate absorption features in dark asteroids spectra, Vilas \& Gaffey (1989) found analogs among CM chondrites. Comparing the whole spectra of G-type asteroids (Tholen taxonomy) and CM chondrites, Burbine (1998) proposed that these asteroids could possibly be the parent bodies of the CM meteorites. 
The 0.7 $\mu$m band has not been found in CI and CR (Cloutis et al., 2011, 2012), but it is commonly seen in CM2 meteorites spectra (Fornasier et al., 1999; Vilas et al., 1994; Burbine, 1998), together with the 3 $\mu$m absorption feature (Jones et al., 1988). These key elements allow us to assume that CM chondrites and the C-class asteroids are linked, and we present a comparison of several spectral parameters of a set of more than 100 CM chondrites from the RELAB database, and the primitive asteroids we studied here. We conducted the same analysis done before for the primitive asteroids to characterize the 0.7 $\mu$m band on the visible spectra of CM chondrites. We found that even if the two populations span the same range of band depth and spectral slope, the band center of meteorites is at longer wavelengths than that of asteroids (Figs.~\ref{metcenslope} and~\ref{metcendepth}). This difference on center positions was described by Burbine (1998) when matching 13 Egeria and 19 Fortuna with CM-type meteorite LEW90500. Cloutis et al. (2011) made an overview of CM chondrites underlining that their 0.7$\mu$m feature is centered between 0.7 and 0.75 $\mu$m  without without any hint of the saponite/olivine complex around 0.65 $\mu$m. He concluded that CM2 meteorites have 
higher abundance of serpentinite-group phyllosilicates, whose absorption bands are centered in the 0.7-0.75 $\mu$m region. Weathering effects, space or terrestrial, can affect spectral reflectance and maybe explain this kind
of mismatch.\\
Moreover, looking at Figures~\ref{metcenslope} and~\ref{metcendepth}, asteroids are more clustered in band depth and spectral slope compared to the CM meteorites. If our meteorite sample is biased, we would have expected it to be more clustered than the asteroids. Nevertheless this behavior is not surprising, because CM meteorites are not an homogeneous group, and several petrologic subtypes have been identified (Cloutis et al., 2011; Rubin et al., 2007). These CM wider range in spectral slope and band depth may also depend on compositional heterogeneities within a single CM, or to spectral changes related to grain size effects or to the presence of different opaques in the meteorites.\\
 Among our set of meteorites, we have the heated and irradiated samples Cloutis et al. (2012) used to illustrate the thermal metamorphism. According to our method of band characterization, we converge towards the same conclusions he made, i.e. decline of the 0.7 $\mu$m  band depth with increasing temperatures and irradiation dose, but we did not notice a shift of the band as a result of these laboratory experiments. \\
From laboratory experiments, different meteorite grain sizes do not produce CM2 band center position shift (Cloutis et al., 2011).
Nevertheless, one has to be careful when using a comparison between
asteroid and laboratory meteorite spectra: the comminution into powder may
not reproduce properly asteroid surface properties. Moreover,  it is possible that the meteorites sample presented here is not fully representative of the primitive asteroids.

[HERE FIG 12 AND 13]

\section{\bf Conclusions}

We have investigated the aqueous alteration process on main belt primitive asteroids on 80 new spectra of primitive asteroids belonging to the C, B, F, P, and G-types, following the Tholen (1984) classification scheme. For a better understanding of the effects of this process on main belt primitive asteroids, we have built a database of 600 visible spectra including the data from our observing campaigns and from available spectroscopic surveys, mainly the SMASS I and II, and the S3OS2 surveys. 
All these spectra have been analyzed in the same manner, and we have characterized the absorption features (mainly the band centered around 0.7 $\mu$m) parameters: band center position, width and depth relative to the continuum, and the spectral slope. Finally we have looked for possible correlations between the aqueous alteration process and the asteroid's taxonomic classes, orbital elements, albedo and sizes. 
The main results coming from the observations here presented, and from the analysis including previously
published visible spectra of main belt primitive asteroids, are the following:
\begin{itemize}
\item We observe absorption features  attributed to hydrated silicates in the new acquired spectra of all the G-type and of 65 \% of the C-type  asteroids. The main feature observed is the 0.7 $\mu$m, that is  associated with $Fe^{2+}\rightarrow Fe^{3+}$ charge transfer absorptions in phyllosilicate minerals, with depth varying between 1\% and 7\% with respect to the continuum.
\item We confirm the strong correlation between the 0.7 $\mu$m band and the 3 $\mu$m band, the deepest feature  associated with hydrated minerals, as 95\% of the asteroids showing the 0.7 $\mu$m band have also the 3 $\mu$m feature.  
\item Considering the sample of 600 primitive asteroids, we find that 4.6\% of P, 7.7\%  of F, 9.8\% of B, 50.5\% of C, and 100\% of the G-type asteroids  have absorption bands in the visible region due to hydrated silicates. Our analysis shows that the aqueous alteration sequence starts from the  P-type objects, practically unaltered, and increases through the P $\rightarrow$ F $\rightarrow$ B $\rightarrow$ C $\rightarrow$ G asteroids, these last being widely aqueous altered, strengthening thus the results previously obtained by Vilas (1994).
\item 45\% of the asteroids belonging to the C-complex (the F, B, C and G classes) have signatures of aqueously altered materials in the visible range.  It must be noted that this percentage represents a lower limit in the number of hydrated asteroids, simply because the absorption features attributed to hydrated silicates are much fainter in the visible range than in the infrared one. Indeed the 3 $\mu$m band, the main absorption feature produced by hydrated silicates, may be present in the spectra of primitive asteroids when no bands are detected in the visible range. All this considered, we estimate that 70\% of the C-complex asteroids might have the 3 $\mu$m signature in the IR range and thus were affected by the aqueous alteration process in the past. 
\item The aqueous alteration process dominates in primitive asteroids located between 2.3 and 3.1 AU, that is at smaller heliocentric distances than previously suggested by Vilas (1994).
\item The percentage of hydrated asteroids is strongly correlated with their size. The aqueous alteration process is less effective for bodies smaller than 50 km, while it dominates in the 50--240 km sized primitive asteroids. According to the proposed heating mechanism, primordial asteroids approaching 50-km diameter have been heated at temperatures high enough to destroy hydrated silicates, while these materials are supposed to survive on larger asteroids. The fact that the percentage of hydrated asteroids decreases with the size is consistent with the hypothesis that the smaller C-complex asteroids we observe nowadays are fragments of hydrated larger asteroids that have been collisionally disrupted and scattered. 
\item No correlation is found between the aqueous alteration process and the asteroids albedo or orbital elements. 
\item Aqueously altered asteroids are the plausible parent bodies of CM2 meteorites. Nevertheless, we see a systematic difference in the 0.7 $\mu$m band center position, the CM2 meteorites having a band centered at longer wavelength (0.71-0.75 $\mu$m) compared to that of hydrated asteroids. Moreover the hydrated asteroids are more clustered in spectral slope and band
depth than the CM meteorites. All these spectral differences may be attributed to different mineral abundances (CM2 meteorites being more serpentine rich than the asteroids), and/or to grain size effects,  or simply to the fact the the CM2 collected on Earth might not be representative of the whole population of aqueous altered asteroids.
\end{itemize}


\vspace{0.3truecm}
{\bf Acknowledgment} \\
We thanks E. Howell and an anonymous referee for their comments and suggestions which helped us to improve this article.
This project was supported by the French Planetology National Program (INSU-PNP).
This research utilizes spectra acquired with the NASA RELAB facility at Brown University.

\bigskip

{\bf References} \\

\noindent

Alexander, C. M. O.D., Bowden, R., Fogel, M. L., Howard, K. T., Herd, C. D. K., Nittler, L. R., 2012. The Provenances of Asteroids, and Their Contributions to the Volatile Inventories of the Terrestrial Planets. Science 337, 721--723

Alvarez-Candal, A., Fornasier, S., Barucci, M. A., de Bergh, C., Merlin, F.
2008. Visible spectroscopy of the new ESO large program on trans-Neptunian objects and Centaurs. Part 1. Astron. Astroph. 487, 741--748

Barucci, M.A., Fulchignoni, M., Lazzarin, M., 1996. Water ice in primitive asteroids? Planet. Space
Sci. 44, 1047--1049 

Barucci, M.A., Doressoundiram, A., Fulchignoni, M., Lazzarin, M., Florczak, M., Angeli, C., Lazzaro, D., 1998. Search for Aqueously Altered Materials on Asteroids. Icarus 132, 388--396 

Bendjoya, P., Cellino, A., Di Martino, M., Saba, L., 2004. Spectroscopic observations of Jupiter Trojans. Icarus 168, 374--384

Binzel R., Rivkin, A. S., Stuart, J. S., Harris, A. W., Bus, S. J., Burbine, T. H., 2004. Observed spectral properties of near-Earth objects: results for population distribution, source regions, and space weathering processes. Icarus 170, 259--294 

Burbine, T.H., Meibon, A., Binzel, R.P., 1996. Meteoritics \& Planetary Sci. 31, 607--620 

Burbine, T.H., 1998. Could G-class asteroids be the parent bodies of the
CM chondrites? Meteoritics \& Planetary Science 33, 253--258

Burns, R. G., 1987. Ferric sulfate on Mars. Proceedings of the XVII Lunar and Planetary Science Conference. J. Geophys. Res. 92, E570--E574

Bus, S. J., Binzel, R. P.,  2002. Phase II of the Small Main-Belt Asteroid Spectroscopic Survey. A Feature-Based Taxonomy. Icarus 158, 146--177

Bus, S. J., Binzel, R.P., 2003. Phase II of the Small Main-Belt Asteroid Spectroscopic Survey. The Observations. Icarus 158, 106--145

Campins, H. et al., 2010. Water ice and organics on the surface of the Asteroid 24
Themis. Nature 464, 1320--1321.

Carvano, J. M., Moth\`e-Diniz, T., Lazzaro, D., 2003. Search for relations among a sample of 460 asteroids with featureless spectra. Icarus 161, 356--382

Cloutis, E.A., Hudon, P., Hiroi, T., Gaffey, M.J., Mann, P., 2011a.
Spectral reflectance properties of carbonaceous chondrites: 2. CM
chondrites. Icarus 216, 309-346

Cloutis, E.A., Hiroi, T., Gaffey, M.J., Alexander, C.M.O'D., Mann, P.,
2011b. Spectral reflectance properties of carbonaceous chondrites: 1. CI
chondrites. Icarus 212, 180-209

Cloutis, E.A., Hudon, P., Hiroi, T., Gaffey, M.J., 2012. Spectral
reflectance properties of carbonaceous chondrites: 3. CR chondrites.
Icarus 217, 389-407

Cloutis, E.A., Hudon, P., Hiroi, T., Gaffey, M.J., 2012. Spectral
reflectance properties of carbonaceous chondrites: 4. Aqueously altered
and thermally metamorphosed meteorites. Icarus 220, 586--617

Cyr, K.E., Sears, W.D., Lunine, J.I., 1998. Distribution and evolution of water ice in the
solar nebula: Implications for Solar System body formation. Icarus 135, 537--548

Dahlgren, M., Lagerkvist, C. I., 1995. A study of Hilda asteroids. I. CCD spectroscopy of Hilda asteroids. Astron. Astrophys. 302, 907--914

de Bergh, C., Boehnhardt, H., Barucci, M. A., Lazzarin, M., Fornasier, S., Romon-Martin, J., Tozzi, G. P., Doressoundiram, A., Dotto, E., 2004. Aqueous altered silicates at the surface of two Plutinos? Astron. Astroph. 416, 791--798

de Leon, J., Lorenzi, V., Ali-Lagoa, V., Licandro, J., Pinilla-Alonso, N., Campins, H., 2013. Additional spectra of asteroid 1996 FG3, backup target of the ESA MarcoPolo-R mission. Astron. Astrophys. 556, A33, 4 pp

De Luise, F., Dotto, E., Fornasier, S., Barucci, A., Pinilla-Alonso, N., Perna, D., Marzari, F., 2010. A peculiar family of Jupiter Trojans: The Eurybates. Icarus 209,  586--590

DeMeo, F.E., Binzel, R.P., Slivan, S.M., Bus, S.J., 2009. An extension of the Bus asteroid
taxanomy into the near-infrared. Icarus 202, 160--180

Dodson-Robinson, S.E., Willacy, K., Bodenheimer, P., Turner, N.J., Beichman, C.A.,
2009. Ice lines, planetesimal composition and solid surface density in the solar
nebula. Icarus, 672--693

Dotto, E., Fornasier, S., Barucci, M. A., Licandro, J., Boehnhardt, H., Hainaut, O., Marzari, F., de Bergh, C., De Luise, F., 2006. The surface composition
of Jupiter Trojans: Visible and Near--Infrared Survey of Dynamical Families. Icarus 183, 420--434 


Emery, J. P., Brown, R. H., 2003. Constraints on the surface composition of Trojan asteroids from near-infrared (0.8-4.0 $\mu$m) spectroscopy. Icarus 164, 104--121

Emery, J. P., Brown, R. H., 2004. The surface composition of Trojan asteroids: constraints set by scattering theory. Icarus 170, 131--152

Emery, J. P., Cruikshank, D. P., Van Cleve, J., 2006. Thermal emission spectroscopy (5.2 38 $\mu$m of three Trojan asteroids with the Spitzer Space Telescope: Detection of fine-grained silicates. Icarus 182, 496--512

Emery, J. P., Burr, D.M., Cruikshank, D. P., 2011. Near-infrared spectroscopy of trojan asteroids: evidence for two compositional groups. AJ 141, 25 (18pp)

Feierberg, M.A., Lebofsky, Tholen, D.J., 1985. The nature of C-class asteroids from 3-$\mu$m spectrophotometry. Icarus, 63, 183--191   

Fornasier, S., Lazzarin, M., Barbieri, C., Barucci, M. A., 1999. Spectroscopic comparison of aqueous altered asteroids with CM2 carbonaceous chondrite meteorites.  Astron. Astrophys.  135, 65--73  

Fornasier, S., Doressoundiram, A., Tozzi, G. P., Barucci, M. A., Boehnhardt, H., 
de Bergh, C., Delsanti A., Davies, J., Dotto, E., 2004a. ESO Large Program on Physical 
Studies of Trans-Neptunian Objects and Centaurs: final results of the visible 
spectroscopic observations. Astron. Astrophys. 421, 353--363

Fornasier, S., Dotto, E., Marzari, F., Barucci, M.A., Boehnhardt, H., Hainaut, O.,
de Bergh, C., 2004b. Visible spectroscopic and photometric survey of L5 Trojans
: investigation of dynamical families. Icarus, 172,  221--232

Fornasier, S., Dotto, E., Hainaut, O., Marzari, F., Boehnhardt, H., De Luise, F., Barucci, M. A. 2007.  Visible spectroscopic and photometric survey of Jupiter Trojans: Final results on dynamical families. Icarus, 190, 622--642

Fornasier, S., Barucci, M. A., de Bergh, C., Alvarez-Candal, A., DeMeo, F., Merlin, F., Perna, D., Guilbert, A., Delsanti, A., Dotto, E., Doressoundiram, A., 2009. Visible spectroscopy of the new ESO large programme on trans-Neptunian objects and Centaurs: final results. Astron. Astrophys. 508, 407--465

Fornasier, S., Clark, B. E., Dotto, E., 2011. Spectroscopic survey of X-type asteroids. Icarus 214, 131--146

Gaffey M. J., Cloutis E. A., Kelley M.S., Reed K., 2002. Mineralogy of Asteroids. In Asteroids III (Bottke W. et al.
editors), Univ. of Arizona Press, Tucson, pp. 183--204

Grimm, R.E., McSween, H.Y., 1989. Water and the thermal evolution of
carbonaceous chondrite parent bodies. Icarus 82, 244--280
Hardorp, J., 1978. The sun among the stars. I- A search for solar spectral analogs. Astron. Astrophys. 63, 383--390 

Hartogh, P., Lis, D. C., Bockelée-Morvan, D., de Val-Borro, M., Biver, N., K\"uppers, M., Emprechtinger, M., Bergin, E. A., Crovisier, J.,  Rengel, M., Moreno, R.,  Szutowicz, S., Blake, G. A., 2012. 
Ocean-like water in the Jupiter-family comet 103P/Hartley 2. Nature 478, 218--220

Herbert, F., Sonnet, C. P.,  Gaffey, M.J., 1991. Protoplanetary
thermal metamorphism: The hypothesis of electromagnetic induction
in the protosolar wind. In The Sun in Time (C. P. Sonnett, M. S.
Giampapa, and M. S. Matthews, Eds.), Univ. of Arizona
Press, Tucson., pp. 710--739. 

Hiroi, T., Zolensky, M.E., Pieter, C.M., Lipschutz, M.E., 1996. Thermal metamorphism of the C, G, B, and F asteroids seen from the 0.7 micron, 3 $\mu$m and UV absorption strengths in comparison with carbonaceous chondrites. Meteoritics \& Planetary Sci. 31, 321--327

Howell, E.S., Rivkin, A.S., Vilas, F, Soderberg, A.M., 2001. Aqueous Alteration in Low Albedo Asteroids. Lunar and Planetary Science Conference 32, 2058  

Howell, E.S., Rivkin, A.S., Vilas, F., Magri, C., Nolan, M. C., Vervack, R. J., Fernandez, Y. R., 2011. Hydrated silicates on main-belt asteroids: Correlation of the 0.7- and 3 $\mu$m absorption bands. EPSC-DPS Joint Meeting 2011, EPSC abstracts, vol. 6, 637

Hsieh, H. H., \& Jewitt, D., 2006. A Population of Comets in the Main Asteroid Belt. Science, 312, 561--563

Hsieh, H. H., Jewitt, D., Fernandez, Y. R., 2009. Albedos of Main-Belt Comets 133P/Elst-Pizarro and 176P/LINEAR. Astroph. J. Letters 694,  L111-L114

Hsieh, H. H., Yang, B., Haghighipour, N., Kaluna, H. M., Fitzsimmons, A., Denneau, L., Novakovic, B., Jedicke, R., Wainscoat, R.J., Armstrong, J. D., and 32 coauthors, 2012. Discovery of Main-belt Comet P/2006 VW139 by Pan-STARRS1. Astroph. J. Letters 748, L15, 7 pp.

Hunt, G. R. \& Ashley R.P.,   1979. Spectra of altered rocks in the visible and near-infrared. Econom. Geol. 74, 1613--1629

Jewitt, D. C., Luu, J. X., 1990. CCD spectra of asteroids. II - The Trojans as spectral analogs of cometary nuclei. Astron. J. 100, 933--944

Johnson, T.V. and Fanale, F.P., 1973. Optical properties of carbonaceous
chondrites and their relationship to asteroids. Journal of Geophysical
Research 78, 8507--8518

Jones, T.D., 1988. An infrared reflectance study of water in outer belt
asteroids : Clues to composition and origin. Ph.D. dissertation

Jones, T. D., Lebofsky, L. A., Lewis, J. S.,  Marley, M. S., 1990. 
The composition and origin of the C, P, and D asteroids: Water
as a tracer of thermal evolution in the outer belt. Icarus, 88, 172--192

King, T.V.V., Clark, R.N., 1989. Spectral characteristics of chlorites and
Mgserpentines using high resolution reflectance spectroscopy. J. Geophys. Res. 94,
13997--14008

King, T.V.V., Clarck, R.N., Calvin, W.M., Sherman, D.M., Brown,
R.H., 1992. Evidence for ammonium-bearing minerals on Ceres. Science 255, 1551--1553 
%

Lazzarin, M., Barucci, M. A., Boehnhardt, H., Tozzi, G. P., de Bergh, C., Dotto, E., 2003. ESO Large Programme on Physical Studies of Trans-Neptunian Objects and Centaurs: Visible Spectroscopy  Astronomical J. 125, 1554--1558

Lazzaro, D., Angeli, C. A., Carvano, J. M., Moth\'e-Diniz, T., Duffard, R., Florczak, M., 2004. S$^{3}$OS$^{2}$: the visible spectroscopic survey of 820 asteroids. Icarus 172, 179--220

Lazzaro, D., Barucci, M. A., Perna, D., Jasmim, F. L., Yoshikawa, M., Carvano, J. M. F., 2013. Rotational spectra of (162173) 1999 JU3, the target of the Hayabusa2 mission. Astron. Astrophys. 549, L2, 4 pp.

Lebofsky L.A. 1980. Infrared reflectance spectra of asteroids: A search for water of hydration.
The Astronomical Journal 85, 573--585

Lebofsky, L.A., Jones, T.D., Owensby, P.D., Feierberg, M.A., Consolmagno, G.J., 1990. The nature of low-albedo asteroids from 3-$\mu$m multi-color photometry. Icarus 83, 16--26

Licandro, J. et al., 2011. 65 Cybele: Detection of small silicate grains, water-ice and
organics. 2011. Astron. Atrophys. 525, A34, 4 pp.

Lunine, 2006, Meteorites and the Early Solar System II, Univ. of Arizona Press, Tucson, pp. 309--319

Mackinnon, I. D. R.,  Rietmeijer, F. J. M. 1987. Mineralogy of chondritic interplanetary dust particles. Rev. Geophys., 25, 1527--1553

McSween Jr., H.Y., Ghosh, A., Grimm, R.E., Wilson, L., Young, E.D., 2002. Thermal
evolution models of asteroid. In: Bottke, W.F., Jr., Cellino, P., Paolicchi, P., Binzel,
R.P. (Eds.), Asteroids III. Univ. of Arizona, pp. 559--571.

Melita, M. D., Licandro, J., Jones, D. C., Williams, I. P. 2008. Physical properties and orbital stability of the Trojan asteroids. Icarus, 195, 686--697

Morbidelli, A, Chambers, J., Lunine, J. I., Petit, J. M., Robert, F., Valsecchi, G. B., Cyr, K. E., 2000. Source regions and time scales for the delivery of water to Earth. Meteoritics \& Planetary Science, 35, 1309--1320

Moroz, L. V., Hiroi, T., Shingareva, T. V., Basilevsky, A. T., Fisenko, A. V., Semjonova, L. F., Pieters, C. M., 2004.
Reflectance Spectra of CM2 Chondrite Mighei Irradiated with Pulsed Laser and Implications for Low-Albedo Asteroids and Martian Moons. Lunar Planet. Sci. Conf. 35, abstract 1279

Perna, D., Dotto, E., Barucci, M. A., Fornasier, S., Alvarez-Candal, A., Gourgeot, F., Brucato, J. R., Rossi, A., 2013. Ultraviolet to near-infrared spectroscopy of the potentially hazardous, low delta-V asteroid (175706) 1996 FG3. Backup target of the sample return mission MarcoPolo-R. Astron. Astrophys.555, A62, 5 pp

Pieters C.M. and McFadden L.A. 1994. Meteorite and asteroid reflectance spectroscopy: Clues
to early solar system processes. Annual Reviews of Earth and Planetary Science 22, 457--497

Rietmeijer, F. J. M, \& MacKinnon, I.D.R., 1985. Layer silicates in a chondritic porous interplanetary dust particle. Journal of Geophys. Res.  90, 149--155

%
Rivkin, A.S., 1997. Observations of Main-Belt Asteroids in the 3-$\mu$m Region. PhD disertation, University of Arizona, Tucson


Rivkin, A. S., Howell, E. S., Vilas, F., Lebofsky L. A., 2002.
Hydrated minerals on asteroids: The astronomical record. In Asteroids III (Bottke W. et al.
editors), Univ. of Arizona Press, Tucson, pp. 235--253

Rivkin, A.S., Emery, J.P., 2010. Detection of ice and organics on an asteroidal surface.
Nature 64, 1322--1323

Rivkin, A.S., 2012. The fraction of hydrated C-complex asteroids in the asteroid belt from SDSS data. Icarus 221, 744--752

Rivkin, A. S., Howell, E. S., Vervack, R. J., Magri, C. Nolan, M. C., Fernandez, Y. R., Cheng, A. F., Barucci, M.A., Michel, P., 2013. The NEO (175706) 1996 FG3 in the 2–4 μm spectral region: Evidence for an aqueously altered surface. Icarus 223, 493--498

Rubin A.E., Trigo-Rodriguez J.M., Huber, H.,  Wasson, J.T., 2007. Progressive aqueous alteration of CM carbonaceous chondrites. Geochim. Cosmoshim. Acta 71, 2361--2382

Sawyer, S.R., 1991, PhD thesis, The University of Texas at Austin. A
hight resolution CCD Spectroscopic Survey of Low Albedo Main Belt Asteroids

Sawyer, S.,1998. $EAR-A-3-RDR-SAWYER-ASTEROID-SPECTRA-V1.2$. NASA Planetary Data System

Shimazu H, Teresawa, T (1995) Electromagnetic induction heating of meteorite parent bodies by the primordial solar wind, Journal of Geoph. Res., 100: 16923--16930

Spearman, C. 1904, The proof and measurements of associations between two
things, AM. J. Psychol., 57, 72


%
Takir D., \& Emery J.P. 2012. Outer Main Belt asteroids: Identification and distribution of four
3-um spectral groups. Icarus 219, 641--654

Taylor S. R., 1992. Solar system evolution: a new perspective. an inquiry into the chemical composition, origin, and
evolution of the solar system. In Solar System evolution: A new Perspective, Cambridge Univ.Press

Tedesco, E.F.,  Noah, P.V.,  Moah, M., Price, S., 2002. The supplemental IRAS
minor planet survey. The Astronomical Journal 123, 10565--10585

Tholen, D.J., 1984. Asteroid taxonomy from cluster analysis of photometry.
Ph.D. dissertation, University of Arizona, Tucson

Tholen, D.J., Barucci, M.A., 1989. Asteroids taxonomy. In: Binzel, R.P.,
Gehrels, T., Matthews, M.S. (Eds.), Asteroids II. Univ. of Arizona Press,
Tucson, pp. 298--315

Townsend, T. E., 1987. Discrimination of iron alteration minerals in visible and near infrared reflectance spectra. J. Geophys. Res. 92, 1441--1454

Vilas, F., \& Smith, B.A., 1985. Reflectance spectrophotometry (about 0.5-1.0 micron) of outer-belt asteroids - Implications for primitive, organic solar system material. Icarus 64, 503--516

Vilas, F.,\& Gaffey, M.J., 1989. Phyllosilicate absorption features in
Main-Belt and Outer-Belt asteroid reflectance spectra. Science 246,
790--792

Vilas, F., McFadden, L.A., 1992. CCD reflectance spectra of selected
asteroids. I. Presentation and data analysis considerations. Icarus 100, 85--94

Vilas, F., Hatch, E.C., Larson, S.M., Sawyer, S.R., Gaffey,
M.J., 1993. Ferric iron in primitive asteroids - A 0.43-$\mu$m absorption feature. Icarus 102, 225--231


Vilas, F., Jarvis, K.S., Gaffey, M.J., 1994. Iron alteration minerals in
the visible and near-infrared spectra of low-albedo asteroids. Icarus 109,
274--283

Vilas, F., 1994. A cheaper, faster, better way to detect water of hydration on Solar System bodies. Icarus, 111, 456--467

Vilas, F., \& Sykes M. W., 1996. Are Low-Albedo Asteroids Thermally Metamorphosed? Icarus 124, 483--489.

Vilas, F., Smith, B.A., McFadden, L.A., Gaffey, M.J., Larson, S.M., Hatch,
E.C., and Jarvis, K.S., 1998. $EAR-A-3-RDR-VILAS-ASTEROID-SPECTRA$. NASA Planetary
Data System

Vilas, F., 2005. Negative Searches for Evidence of Aqueous Alteration on Asteroid Surfaces. In: Mackwell, S., Stansbery, E. (Eds.), Lunar Planet. Sci. 36, abstract no. 2033.

Vilas, F., 2008. Spectral Characteristics of Hayabusa 2 Near-Earth Asteroid Targets 162173 1999 JU3 and 2001 QC34. AJ 135, 1101--1105

Walsh, K.J., Morbidelli, A., Raymond, S.N., O’Brien, D.P., Mandell, A.M., 2011. A low
mass for Mar. from Jupiter’s early gas-driven migration. Nature 475, 206--209

Xu, S. Binzel, R. P., Burbine, T. H., Bus, S. J. 1995. Small main-belt asteroid spectroscopic survey: Initial results. Icarus 155, 1--35


\newpage

{\bf Tables}

{\scriptsize
       \begin{center}
     \begin{longtable} {|l|c|c|c|c|c|c|c|}  
\caption[]{Observational circumstances for the observed asteroids. Tx is the Tholen (1984) taxonomic class.} 
        \label{tab1} \\
\hline \multicolumn{1}{|c|} {\textbf{Asteroid }} & \multicolumn{1}{c|}
{\textbf{Date}} & \multicolumn{1}{c|} {\textbf{UT$_{start}$}} & \multicolumn{1}{c|}
{\textbf{m$_v$}} & \multicolumn{1}{c|} {\textbf{T$_{exp}$}} &     \multicolumn{1}{c|} {\textbf{Airm.}} & \multicolumn{1}{c|} {\textbf{Solar Analog (airm.)}} & \multicolumn{1}{c|}
{\textbf{Tx}}   \\  
 \multicolumn{1}{|c|}{\textbf{}}  & \multicolumn{1}{c|}{\textbf{}}
 & \multicolumn{1}{c|} {\textbf{(hh:mm)}} & \multicolumn{1}{c|}{\textbf{}}
 & \multicolumn{1}{c|} {\textbf{(s)}} &     \multicolumn{1}{c|}{\textbf{}}  & \multicolumn{1}{c|}{\textbf{}} & \multicolumn{1}{c|}{\textbf{}}
  \\  \hline  \hline
\endfirsthead
\multicolumn{8}{c}%
{{\bfseries \tablename\ \thetable{} -- continued from previous page}} \\ \hline 
\endfoot
\hline \multicolumn{1}{|c|} {\textbf{Asteroid    }} & \multicolumn{1}{c|}
{\textbf{Date}} & \multicolumn{1}{c|} {\textbf{UT$_{start}$}} & \multicolumn{1}{c|}
{\textbf{m$_v$}} & \multicolumn{1}{c|} {\textbf{T$_{exp}$}} &     \multicolumn{1}{c|} {\textbf{Airm.}} & \multicolumn{1}{c|} {\textbf{Solar Analog (airm.)}} & \multicolumn{1}{c|}
{\textbf{Tx}}   \\  
\multicolumn{1}{|c|}{\textbf{}}  & \multicolumn{1}{c|}{\textbf{}}
 & \multicolumn{1}{c|} {\textbf{(hh:mm)}} & \multicolumn{1}{c|}{\textbf{}}
 & \multicolumn{1}{c|} {\textbf{(s)}} &     \multicolumn{1}{c|}{\textbf{}}  & \multicolumn{1}{c|}{\textbf{}} & \multicolumn{1}{c|}{\textbf{}} \\
\hline 
\hline 
\endhead
\hline \multicolumn{8}{r}{{Continued on next page}} \\ 
\endfoot
\hline \hline
\endlastfoot
10 Hygiea & 15 Mar. & 03:01 & 10.2 & 60 & 1.30 & HD44594 (1.32) & C \\
13 Egeria &  04 Nov. & 06:43 & 10.0 & 60 & 1.62 & Hyades64 (1.45) & G \\
31 Euphrosyne &  04 Nov. & 03:10 & 12.5 & 360 & 1.25 & HD20630 (1.18) & C \\
36 Atalante & 16 Mar. & 06:27 & 14.2 & 720 & 1.02 & HD144585 (1.16) & C\\
38 Leda &  04 Nov. & 02:22 & 13.5 &720 & 1.41 & Hyades64 (1.45)& C\\
47 Aglaja & 04 Nov.  & 05:13 & 11.8 & 480 & 1.59 & Hyades64 (1.45)& C \\
48 Doris & 04 Nov. & 01:21 & 13.0 & 480 & 1.70 & Hyades64 (1.45)& CG \\
54 Alexandra & 15 Mar. & 00:57 & 13.2 & 540 & 1.61 & HD44594 (1.32) & C \\
56 Melete & 15 Mar. & 09:56 & 12.6 & 300 & 1.02 & HD44594 (1.15) & P  \\
58 Concordia &  05 Nov. & 02:07 & 13.9 & 900 & 1.21 & Hyades64 (1.47)& C  \\
66 Maja & 16 Mar. & 05:57 & 14.2 & 720 & 1.08 & HD144585 (1.16)& C  \\
78 Diana &15 Mar. & 08:33 & 11.8 & 180 & 1.07 & HD76151 (1.15) & C  \\
81 Terpsicore &04 Nov.  & 03:19  & 12.2 & 300 & 1.30 & Hyades64 (1.45)& C  \\
84 Klio & 15 Mar. & 09:40 & 13.9 & 660 & 1.01 &  HR6060 (1.10) & G \\
85 Io &  04 Nov. & 03:28 & 11 & 120 & 1.25 & Hyades64 (1.45) & FC  \\
86 Semele &  15 Mar. & 05:36 & 13.6 & 600 & 1.39 & HD44594 (1.32)& C \\
90 Antiope & 17 Mar. & 08:33  & 13.7 & 540 &1.05 & HR6060 (1.10)& C \\
93 Minerva & 04 Nov. & 00:34 & 12.7 & 420 & 1.38 & Hyades64 (1.45)& CU  \\
95 Arethusa & 16 Mar. & 06:52 & 13.5 & 480 & 1.01 & HD144585 (1.16)& C \\
107 Camilla &  04 Nov. & 08:05 & 12.6 & 360 & 1.30 & Hyades64 (1.45)& C  \\
120 Lachesis &  15 Mar. & 08:53 & 12.7 & 300 & 1.02 & HR6060 (1.10)& C \\
121 Hermione &  15 Mar. & 06:18 & 13.0 & 420 & 1.26&  HD44594 (1.29)& C \\
134 Sophrosyne & 16 Mar. & 07:06 & 13.6 & 540 & 1.09 & HD144585 (1.16)& C \\
139 Juewa & 05 Nov. & 02:48 & 13.2 & 600 & 1.23 & Hyades64 (1.47) & CP  \\
140 Siwa &  15 Mar. & 06:01 & 12.5 & 300 & 1.22 & HD44594 (1.32) & P \\
140 Siwa & 16 Mar. & 05:08 & 12.5 & 300 & 1.20 & HD44594 (1.29)& P\\
142 Polana & 16 Mar. & 00:04 & 14.7 & 780 & 1.61 & HD44594 (1.29)& F\\
143 Adria &  15 Mar. & 05:14 & 13.0 & 420 & 1.35 & HD44594 (1.32)& C\\
145 Adeona &  04 Nov. & 05:45  & 11.6 & 240 & 1.24 & Hyades64 (1.45)& C \\
148 Gallia & 04 Nov. & 00:18 & 13.2 & 480 & 1.46 &  Hyades64 (1.45) & GU \\
150 Nuwa &  04 Nov. & 07:35 & 12.3 & 720 & 1.67 & Hyades64 (1.45) & CX\\
153 Hilda & 16 Mar. & 09:41 & 13.6 & 540 & 1.02 & HD144585 (1.16)& P \\
156 Xanthippe &  15 Mar. & 03:14 & 12.2 & 300 & 1.14 & HD44594 (1.29) & C \\
175 Andromache & 05 Nov. & 06:45 & 12.7 & 420 & 1.73 & Hyades64 (1.47) & C \\
176 Iduna &  15 Mar. & 07:00 & 13.3 & 420 & 1.11 & HR6060 (1.10)& G \\
194 Prokne &  04 Nov. & 07:24  & 11.2 & 120 & 1.20 & Hyades64 (1.45)& C \\
205 Martha & 05 Nov. & 02:28 & 13.8 & 840 & 1.29 & Hyades64 (1.47)& C  \\
206 Hersilia & 16 Mar. & 02:26 & 13.2 & 420 & 1.62 &Hyades64 (1.78)& C \\
209 Dido & 16 Mar. & 02:58 & 13.2 & 420 & 1.48 & HD144585 (1.16) & C \\
213 Lilaea & 16 Mar. & 07:52 & 12.5 & 240 & 1.26& HD44594 (1.29) & F\\
238 Hypatia & 05 Nov. & 07:09 & 12.8 & 420 & 1.27 & Hyades64 (1.47)& C\\
240 Vanadis & 16 Mar. & 00:50 & 13.2 & 420 & 1.58 & Hyades64 (1.78) & C \\
259 Aletheia &  04 Nov. & 06:58 & 12.7 & 420 & 1.58 & Hyades64 (1.45)& CP \\
313 Chaldea &  15 Mar. & 00:03 & 12.5 & 180 & 1.33 &HD44594 (1.29)& C \\
329 Svea &  04 Nov. & 07:10 & 13.3 & 540 & 1.24 & Hyades64 (1.45)& C \\
331 Etheridgea & 05 Nov. & 03:37 & 13.8 & 900 & 1.27 &  Hyades64 (1.47)& CX\\
334 Chicago & 05 Nov. & 01:01 & 14.2 & 1200 & 1.21 & HD20630 (1.20)& C \\
342 Endymion &  04 Nov. & 05:32 & 13 & 480 & 1.61 &  Hyades64 (1.45)& C\\
356 Liguria & 16 Mar. & 06:14  & 13.0 & 420 & 1.04 & HR6060 (1.07) & C\\
360 Carlova & 05 Nov. & 00:30 & 14.4 & 1320 & 1.40 & HD20630 (1.20)& C \\
372 Palma & 04 Nov. &  02:05 & 13.2 & 480 & 1.49 & Hyades64 (1.45) & BFC \\
381 Myrrha & 17 Mar. & 09:45 & 13.9 & 720 & 1.14 & HR6060 (1.07)& C \\
386 Siegena & 16 Mar. & 09:55 & 13.4 & 480 &  1.15 & HD144585 (1.16)& C  \\
393 Lampetia &  15 Mar. & 05:52 & 12.9 & 480 & 1.14 & HR6060  (1.10)& C \\
395 Delia & 17 Mar. & 08:10  & 14.9 & 1020 & 1.02 & HR6060 (1.07)& C \\
405 Thia & 05 Nov. & 01:28 & 13.9 & 1020 & 1.40 & Hyades64 (1.47)& C \\
407 Arachne & 17 Mar. & 04:39 & 13.2 & 420 & 1.82 & Hyades64 (1.80) & C \\
409 Aspasia &  04 Nov. & 08:59  & 12.6 & 300 & 1.40 & Hyades64 (1.45)& CX\\
414 Liriope & 05 Nov. & 05:11  & 14.3 & 1200 & 1.24 & HD20630 (1.20) & C \\
419 Aurelia &  15 Mar. & 00:20  & 13.5 & 660 & 1.44 & HD44594 (1.32)& F \\
449 Hamburga &  15 Mar. & 07:16  & 12.4 & 240 & 1.18 & HR6060 (1.10) & C \\
476 Hedwig & 16 Mar. & 03:45 & 12.5 & 300 & 1.09 & HR6060  (1.10) & P  \\
489 Comacina & 05 Nov. & 07:49 & 13.9 & 900 & 1.41 & Hyades64 (1.47) & C  \\
511 Davida & 16 Mar. & 09:32 & 12.3 & 270 & 1.09 & HR6060 (1.07) & C  \\
559 Nanon & 16 Mar. & 07:29 & 13.6 & 840 & 1.01 & HR6060 (1.07) & C \\
583 Klotilde & 17 Mar. & 09:21 & 14.4 & 840 & 1.01& HR6060 (1.07)& C \\
585 Bilkis  & 05 Nov. & 04:53 & 13.6 & 780 & 1.27 & HD20630 (1.20)& C  \\
602 Marianna & 16 Mar. & 05:37 & 14.0 & 720 & 1.03 &HR6060 (1.10)& C \\
654 Zelinda &   15 Mar. & 09:10 & 12.2 & 180 & 1.04 & HD76151 (1.15)& C \\
702 Alauda &  15 Mar. & 02:51 & 12.5 & 300 & 1.25 &   HD76151 (1.15) & C \\
704 Interamnia &  15 Mar. & 08:45  & 11.7 & 180 & 1.12 & HR6060 (1.10)& F \\
748 Simeisa & 16 Mar. & 04:49 & 14.2 & 780 & 1.15 & HD44594 (1.29) & P \\
786 Bredichina & 16 Mar. & 07:42 & 13.3 & 480 & 1.16 & HR6060 (1.07) & C  \\
790 Pretoria & 16 Mar. & 23:56 & 14.6 & 900 & 1.36 & HD44594 (1.29) & P \\
814 Tauris &  04 Nov. & 05:54 & 11.9 & 240 & 1.11 &  HD20630 (1.18)& C \\
868 Lova & 05 Nov. & 04:39 & 13.0 & 540 & 1.22 & HD20630 (1.20) & C \\
977 Philippa & 17 Mar. & 07:30 & 14.6 & 840 & 1.22&  HD144585 (1.07)& C \\
1015 Christa & 16 Mar. & 04:29 & 13.9 & 780 & 1.35 & HD44594 (1.62) & C \\
1021 Flammario &  04 Nov. & 05:22 & 11.7 & 240 & 1.50 & Hyades64 (1.45) & F\\
1268 Lybia & 17 Mar. & 05:08 & 14.6 & 780 & 1.70 & HD89010 (2.00) & P \\
1330 Spiridonia & 17 Mar. & 06:48 & 14.3 & 780 & 1.66 & HD44594 (1.62)& P   \\
\hline
\end{longtable}
\end{center}
}
\begin{list}{}{}
\item 
\end{list}

{\scriptsize
\begin{center}
\begin{longtable}{|l|c|c|c|c|c|c|c|c|c|c|c|}
\caption[]{Characterization of the absorption bands, when present, and of the spectral slope of the observed primitive asteroids. H indicates the hydrated state. The albedo comes from WISE data when available, or from IRAS (indicated with the symbol $^I$). These asteroids make part of the 0.7 $\mu$m band statistical study we also present later in this paper. $^{*}$ Hydrated asteroids without the 0.7 $\mu$m absorption band. They won't be used for
the study of this specific band but only to know the percentage of
hydrated primitive asteroids. $^{+}$ We did not include these data in the statistical study for several reasons: better data or detected
absorption band in another database (see Table~\ref{tab_all}). 
\textbf{References.} (1) Fornasier et al., 1999; (2) this
study. $^{a}$: observed at Asiago on March 1997; $^{b}$: observed at
ESO December 1997; $^{c}$: observed at Asiago on June 1997; $^{d}$: observed at ESO on
15 March 1999; $^{e}$: observed at ESO on 16 March 1999.}
\label{tab2} \\
\hline \multicolumn{1}{|c|} {\textbf{Asteroid}} & \multicolumn{1}{c|}
{\textbf{a}} &\multicolumn{1}{c|}
{\textbf{Alb.}} & \multicolumn{1}{c|} {\textbf{Diam.}} & \multicolumn{2}{c|}
{\textbf{Taxon.}} & \multicolumn{1}{c|} {\textbf{H}} &     \multicolumn{1}{c|} {\textbf{Center}} & \multicolumn{1}{c|} {\textbf{Depth}} & \multicolumn{1}{c|}{\textbf{Width}} & \multicolumn{1}{c|} {\textbf{Slope}} & \multicolumn{1}{c|} {\textbf{R}}    \\  
\multicolumn{1}{|c|}{\textbf{}}  & \multicolumn{1}{c|}{\textbf{(AU)}}
 & \multicolumn{1}{c|} {\textbf{(\%)}} & \multicolumn{1}{c|}{\textbf{(km)}}
 & \multicolumn{1}{c|} {\textbf{(Th)}} &     \multicolumn{1}{c|}{\textbf{(B)}}  & \multicolumn{1}{c|}{\textbf{}} & \multicolumn{1}{c|}{\textbf{(\AA)}} & \multicolumn{1}{c|}{\textbf{(\%)}} & \multicolumn{1}{c|}{\textbf{(\AA)}} & \multicolumn{1}{c|}{\textbf{\%/($10^{3}$\AA)}} & \\ \hline
\endfirsthead
\multicolumn{12}{c}%
{{\bfseries \tablename\ \thetable{} -- continued from previous page}} \\ \hline 
\endfoot
\hline 
\hline \multicolumn{1}{|c|} {\textbf{Asteroid}} & \multicolumn{1}{c|}
{\textbf{a}} &\multicolumn{1}{c|}
{\textbf{Alb.}} & \multicolumn{1}{c|} {\textbf{Diam.}} & \multicolumn{2}{c|}
{\textbf{Taxon.}} & \multicolumn{1}{c|} {\textbf{H}} &     \multicolumn{1}{c|} {\textbf{Center}} & \multicolumn{1}{c|} {\textbf{Depth}} & \multicolumn{1}{c|}{\textbf{Width}} & \multicolumn{1}{c|} {\textbf{Slope}} & \multicolumn{1}{c|} {\textbf{R}}    \\  
\multicolumn{1}{|c|}{\textbf{}}  & \multicolumn{1}{c|}{\textbf{(AU)}}
 & \multicolumn{1}{c|} {\textbf{(\%)}} & \multicolumn{1}{c|}{\textbf{(km)}}
 & \multicolumn{1}{c|} {\textbf{(Th)}} &     \multicolumn{1}{c|}{\textbf{(B)}}  & \multicolumn{1}{c|}{\textbf{}} & \multicolumn{1}{c|}{\textbf{(\AA)}} & \multicolumn{1}{c|}{\textbf{(\%)}} & \multicolumn{1}{c|}{\textbf{(\AA)}} & \multicolumn{1}{c|}{\textbf{\%/($10^{3}$\AA)}} & \\ \hline
\hline
\endhead
\hline \multicolumn{12}{r}{{Continued on next page}} \\ 
\endfoot
\hline \hline
\endlastfoot
1 Ceres$^*$ & 2.768 & 11.3$\pm$0.5$^I$ & 848.4$\pm$19.7 & G & C & Y & 8960$\pm$51 & 5.0$\pm$0.1 & 8214-9439 & -0.54$\pm$0.50 & 1 \\
10 Hygiea & 3.137 & 5.8$\pm$0.5 & 453.2$\pm$19.2 & C & C & N & - &-& - & -1.72$\pm$0.50 & 1 \\
 & & & & & & Y & 6631$\pm$69 & 1.0$\pm$0.1 & 5278-7747 & 0.02$\pm$0.50 & 2 \\
13 Egeria & 2.576 & 6.9$\pm$2.2 & 227.0$\pm$25.9 & G & Ch & Y & 4372$\pm$10 & 1.9$\pm$0.1 & 4264-4511 & -1.55$\pm$0.50 & 2 \\
 & & & & & & & 6815$\pm$69 & 2.4$\pm$0.1 & 5334-7899 & & \\
19 Fortuna & 2.442 & 5.0$\pm$2.0 & 223.0$\pm$43.6 & G & Ch & Y & 7147$\pm$52 & 5.5$\pm$0.2 & 5857-8504 & -1.80$\pm$0.52 & 1$^a$ \\
 & & & & & & Y & 6977$\pm$27 & 5.0$\pm$0.2 & 5228-8570 & -1.19$\pm$0.51 & \textbf{1$^b$}  \\
24 Themis & 3.138 & 6.4$\pm$1.6 & 202.3$\pm$6.1 & C & B & Y & 6722$\pm$39 & 3.5$\pm$0.1 & 5508-8061 & 0.13$\pm$0.51 & 1 \\
31 Euphrosyne$^*$ & 3.153 & 5.4$\pm$0.5 & 255.9$\pm$11.5 & C & Cb & Y & 4416$\pm$16 & 2.3$\pm$0.1 & 4234-4630 & 1.27$\pm$0.50 & 2 \\
34 Circe & 2.687 & 5.4$\pm$1.3 & 113.2$\pm$2.9 & C & Ch & Y & 6971$\pm$10 & 1.7$\pm$0.2 & 5454-8473 & 0.37$\pm$0.51 & 1 \\
36 Atalante & 2.749 & 6.9$\pm$1.2 & 103.0$\pm$11.5 & C & - & Y & 6795$\pm$45 & 3.4$\pm$0.1 & 5498-8368 & -0.45$\pm$0.51 & 2 \\
38 Leda & 2.739 & 6.2$\pm$1.6 & 116.0$\pm$15.5 & C & Cgh & Y & 6869$\pm$16 & 2.6$\pm$0.1 & 5558-8073 & 1.59$\pm$0.51 & 1 \\
 & & & & & & Y & 6831$\pm$37 & 2.6$\pm$0.1 & 5232-8189 & 0.05$\pm$0.50 & \textbf{2} \\
41 Daphne & 2.761 & 8.3$\pm$1.2$^I$& 174.0$\pm$11.7 & C & Ch & Y & 6925$\pm$10 & 3.3$\pm$0.2 & 5324-8483 & -0.13$\pm$0.50 & 1 \\
45 Eugenia & 2.720 & 4.6$\pm$0.5 & 206.1$\pm$6.2 & C & C & N & - &
-& - & 0.81$\pm$0.50 & 1 \\
47 Aglaja$^*$ & 2.880 & 6.7$\pm$0.9 & 138.0$\pm$11.1 & C & B & Y & 5991$\pm$22 & 0.9$\pm$0.1 & 5378-6429 & -0.48$\pm$0.50 & 2 \\
48 Doris$^+$ & 3.111 & 6.2$\pm$1.4 & 223.4$\pm$4.2 & C & Ch & Y & 6810$\pm$131 & 2.3$\pm$0.1 & 5338-7820 & -1.42$\pm$0.50 & 2\\
 & & & & & & & 8227$\pm$34 & 4.3$\pm$0.3 & 7785-8497 & & \\
51 Nemausa & 2.365 & 10.0$\pm$2.6 & 142.6$\pm$12.5 & C & Ch & Y & 7178$\pm$71 & 6.3$\pm$0.2 & 5439-8432 & 1.67$\pm$0.52 & $1^a$ \\
 & & & & & & Y & 7064$\pm$18 & 4.3$\pm$0.1 & 5754-8366 & 0.80$\pm$0.51 & \textbf{1$^c$} \\								 
54 Alexandra & 2.710 & 4.9$\pm$0.8 & 142.0$\pm$14.8 & C & C & Y & 6778$\pm$71 & 3.9$\pm$0.1 & 5228-8138 & 0.92$\pm$0.51 & 2 \\
56 Melete & 2.597 & 5.0$\pm$0.5 & 129.1$\pm$4.4 & P & Xk & N &- &- &- & 5.49$\pm$0.50 & 2 \\
58 Concordia & 2.699 & 5.9$\pm$0.5 & 92.3$\pm$1.5 & C & Ch & Y & 7067$\pm$60 & 2.4$\pm$0.1 & 5324-8377 & -0.61$\pm$0.50 & 2 \\
65 Cybele$^+$ & 3.426 & 7.1$\pm$0.3$^I$ & 237.3$\pm$4.2 & P & Xc & N &- &- &- & 3.05$\pm$0.50 & 1 \\
66 Maja & 2.647 & 6.2$\pm$1.0$^I$ & 71.8$\pm$5.3 & C & Ch & Y & 6762$\pm$50 & 3.8$\pm$0.1 & 5512-8307 & -0.34$\pm$0.51 & 2 \\
70 Panopaea & 2.615 & 4.0$\pm$0.9 & 139.0$\pm$3.9 & C & Ch & Y & 6837$\pm$45 & 2.4$\pm$0.2 & 5505-8281 & 1.1$\pm$0.51 & 1 \\
74 Galatea & 2.777 & 4.3$\pm$0.2$^I$ & 118.7$\pm$2.8 & C & C & N &- &- &- & 0.06$\pm$0.50 & 1 \\
78 Diana & 2.620 & 7. 1$\pm$0.3$^I$ & 120.6$\pm$2.7 & C & Ch & Y & 6873$\pm$20 & 2.4$\pm$0.1 & 5467-8354 & -0.17$\pm$0.50 & 2 \\
81 Terpsichore & 2.856 & 3.4$\pm$0.3 & 121.6$\pm$3.2 & C & Cb & N &- &- &- & 1.24$\pm$0.50 & 2 \\
84 Klio & 2.362 & 5.3$\pm$1.7 & 79.0$\pm$4.9 & G & Ch & Y & 7025$\pm$25 & 2.9$\pm$0.1 & 5628-8347 & 1.71$\pm$0.50 & 2 \\
85 Io$^+$ & 2.655 & 6.3$\pm$2.5 & 163.0$\pm$18.6 & F & B & N & -& -& -& -0.36$\pm$0.50 & 2 \\
86 Semele & 3.112 & 5.1$\pm$0.5 & 115.5$\pm$2.5 & C & - & N & -& -& -& 0.59$\pm$0.50 & 2 \\
90 Antiope$^+$ & 3.164 & 5.9$\pm$1.4 & 121.1$\pm$3.5 & C & C & Y & 6934$\pm$36 & 0.8$\pm$0.1 & 5538-8212 & -0.08$\pm$0.50 & 2 \\
 & & & & & & & 8609$\pm$10 & 1.3$\pm$0.1 & 8234-8946 & & \\
93 Minerva$^*$ & 2.755 & 7.3$\pm$0.4$^I$ & 141.6$\pm$4.0 & C & C & Y &  4284$\pm$16 & 2.3$\pm$0.1 & 4093-4559 & 0.92$\pm$0.50 & 2 \\
 & & & & & & & 8275$\pm$10 & 1.8$\pm$0.1 & 7890-8656 & & \\
95 Arethusa & 3.069 & 5.7$\pm$2.5 & 150.2$\pm$7.1 & C & Ch & Y & 6847$\pm$19 & 3.8$\pm$0.1 & 5505-8323 & -0.16$\pm$0.51 & 2 \\
104 Klymene & 3.150 & 5.5$\pm$0.6 & 125.8$\pm$2.6 & C & Ch & Y & 6855$\pm$53 & 1.8$\pm$0.1 & 5432-8400 & 0.32$\pm$0.50 & 1 \\
105 Artemis & 2.373 & 4.7$\pm$1.2 & 119.0$\pm$17.3 & C & Ch & Y & 6944$\pm$32 & 3.2$\pm$0.1 & 5420-8324 & -0.74$\pm$0.50 & 1 \\
107 Camilla$^+$ & 3.491 & 5.4$\pm$1.1 & 219.4$\pm$5.9 & C & X & N & -& -& -& 1.73$\pm$0.50 & 2 \\
120 Lachesis & 3.118 & 5.2$\pm$1.2 & 164.6$\pm$5.2 & C & C & N & -& -& -& 1.52$\pm$0.50 & 2 \\
121 Hermione & 3.450 & 7.7$\pm$1.0 & 165.0$\pm$4.5 & C & Ch & Y & 6891$\pm$32 & 2.5$\pm$0.1 & 5506-8515 & -0.03$\pm$0.50 & 2 \\
128 Nemesis$^+$ & 2.750 & 5.0$\pm$1.3 & 188.0$\pm$9.0 & C & C & N & -& -& -& 0.94$\pm$0.50 & 1 \\
130 Elektra & 3.124 & 7.1$\pm$1.1 & 198.9$\pm$4.1 & G & Ch & Y & 7070$\pm$10 & 3.2$\pm$0.1 & 5719-8535 & 0.88$\pm$0.51 & 1 \\
134 Sophrosyne & 2.563 & 4.4$\pm$1.6 & 112.2$\pm$10.8 & C & Ch & Y & 6702$\pm$47 & 3.0$\pm$0.1 & 5377-8344 & 0.96$\pm$0.50 & 2 \\
137 Meliboea & 3.119 & 5.1$\pm$1.1 & 144.0$\pm$11.3 & C & - & Y & 6880$\pm$10 & 2.9$\pm$0.1 & 5408-8355 & 0.44$\pm$0.50 & 1 \\
139 Juewa$^*$ & 2.781 & 4.5$\pm$2.3 & 164.0$\pm$25.2 & C & X & Y & 4348$\pm$11 & 1.2$\pm$0.1 & 4230-4519 & 1.57$\pm$0.50 & 2 \\
 & & & & & & & 8741$\pm$40 & 1.7$\pm$0.1 & 8426-9177 & & \\
140 Siwa & 2.733 & 6.8$\pm$0.4$^I$ & 109.8$\pm$3.0 & P & Xc & N & -& -& -& 4.66$\pm$0.50 & \textbf{2$^d$} \\
 & & & & & & N & -& -& -& 3.89$\pm$0.50 & 2$^e$ \\
142 Polana & 2.418 & 4.5$\pm$0.6 & 56.6$\pm$1.4 & F & B & N & -& -& -& 0.17$\pm$0.51 & 2 \\
143 Adria & 2.761 & 5.3$\pm$1.5 & 86.3$\pm$2.3 & C & Xc & N & -& -& -& 3.06$\pm$0.50 & 2 \\
144 Vibilia & 2.655 & 6.0$\pm$0.2$^I$ & 142.4$\pm$2.6 & C & Ch & Y & 6937$\pm$30 & 2.9$\pm$0.1 & 5324-8493 & 0.02$\pm$0.50 & 1 \\
145 Adeona & 2.674 & 4.3$\pm$1.4 & 151.0$\pm$14.2 & C & Ch & Y & 7384$\pm$117 & 2.7$\pm$0.1 & 5494-8429 & -1.30$\pm$0.51 & 1 \\ 
 & & & & & & Y & 4352$\pm$12 & 2.1$\pm$0.1 & 4238-4489 & -0.88$\pm$0.50 & \textbf{2}  \\
 & & & & & & & 6976$\pm$57 & 3.6$\pm$0.1 & 5316-8260 & & \\      
146 Lucina & 2.718 & 5.3$\pm$1.0 & 131.8$\pm$4.8 & C & Ch & Y & 6870$\pm$31 & 2.5$\pm$0.2 & 5601-8248 & 0.07$\pm$0.51 & 1 \\
150 Nuwa$^*$ & 2.983 & 4.8$\pm$1.0 & 137.2$\pm$3.4 & C & Cb & Y & 4304$\pm$10 & 1.9$\pm$0.1 & 4135-4480 & 0.79$\pm$0.50 & 2 \\
 & & & & & & & 8249$\pm$40 & 1.6$\pm$0.1 & 7785-8581 & & \\
153 Hilda & 3.973 & 6.2$\pm$0.2$^I$ & 170.6$\pm$3.3 & P & X & N & -& -& -& 4.41$\pm$0.50 & 2 \\
156 Xanthippe & 2.728 & 5.0$\pm$1.2 & 110.7$\pm$2.2 & C & Ch & Y & 6998$\pm$10 & 2.8$\pm$0.1 & 5549-8452 & 1.16$\pm$0.50 & 2 \\
175  Andromache$^*$ & 3.183 & 6.3$\pm$0.9 & 115.3$\pm$1.0 & C & Cg & Y & 4372$\pm$10 & 1.5$\pm$0.1 & 4261-4493 & -0.24$\pm$0.50 & 2 \\
& & & & & & & 8374$\pm$33 & 1.6$\pm$0.1 & 7948-8844 & & \\
176 Iduna & 3.187 & 8.2$\pm$1.2 & 122.2$\pm$2.7 & G & Ch & Y & 7047$\pm$56 & 4.8$\pm$0.2 & 5453-8342 & -2.02$\pm$0.51 & 2 \\
 & & & & & & & 8586$\pm$20 & 1.8$\pm$0.1 & 8358-8916 & & \\
185 Eunike & 2.740 & 6.4$\pm$1.5 & 157.0$\pm$8.2 & C & C & N & -& -& -& -0.73$\pm$0.51 & 1 \\
190 Ismene & 4.001 &  - &  - & P & X & N & -& -& -& 1.93$\pm$0.50 & 1 \\
194 Prokne & 2.618 & 5.2$\pm$1.5 & 169.0$\pm$14.5 & C & C & Y & 6903$\pm$18 & 2.7$\pm$0.1 & 5303-8379 & 0.16$\pm$0.50 & 2 \\
200 Dynamene & 2.736 & 5.2$\pm$0.5 & 130.5$\pm$2.9 & C & Ch & Y & 7000$\pm$17 & 2.2$\pm$0.3 & 5408-8527 & 1.67$\pm$0.51 & 1 \\
 & & & & & & & 8265$\pm$14 & 1.9$\pm$0.1 & 7978-8679 & & \\
205 Martha & 2.777 & 5.4$\pm$0.9 & 81.5$\pm$0.8 & C & Ch & Y & 7016$\pm$52 & 1.5$\pm$0.1 & 5334-8410 & 0.66$\pm$0.50 & 2 \\
206 Hersilia & 2.740 & 5.5$\pm$0.8 & 104.6$\pm$3.4 & C & C & N & -& -& -& 1.28$\pm$0.50 & 2 \\
209 Dido & 3.145 & 5.8$\pm$1.9 & 124.3$\pm$3.7 & C & Xc & N & -& -& -& 0.91$\pm$0.50 & 2 \\
211 Isolda & 3.041 & 6.0$\pm$1.8 & 143.0$\pm$21.6 & C & Ch & Y & 7267$\pm$39 & 2.1$\pm$0.1 & 5409-8410 & 0.08$\pm$0.51 & 1 \\
213 Lilaea & 2.752 & 9.0$\pm$0.6$^I$ & 83.0$\pm$2.6 & F & B & N & -& -& -& -0.01$\pm$0.50 & 2 \\
238 Hypatia & 2.908 & 4.4$\pm$0.6 & 146.5$\pm$8.7 & C & Ch & N & -& -& -& 2.54$\pm$0.51 & 1 \\
 & & & & & & Y & 6719$\pm$19 & 1.3$\pm$0.1 & 5453-8054 & 0.38$\pm$0.50 & \textbf{2}  \\	 
240 Vanadis & 2.665 & 5.3$\pm$1.0 & 91.4$\pm$2.6 & C & C & Y & 6894$\pm$40 & 2.8$\pm$0.1 & 5463-8473 & 1.71$\pm$0.50 & 2 \\
259 Aletheia$^*$ & 3.133 & 4.2$\pm$0.5 & 182.9$\pm$3.5 & C & X & Y & 4359$\pm$16 & 1.3$\pm$0.1 & 42445-4515 & 1.10$\pm$0.50 & 2 \\
 & & & & & & & 8308$\pm$56 & 1.4$\pm$0.1 & 7754-8704 & & \\
304 Olga & 2.405 & 4.4$\pm$0.8 & 68.9$\pm$2.3 & C & Xc & N & -& -& -& 2.04$\pm$0.50 & 1 \\
313 Chaldaea & 2.375 & 5.3$\pm$1.3 & 96.0$\pm$7.8 & C & - & Y &  7091$\pm$59 & 3.0$\pm$0.1 & 5387-8323 & -0.84$\pm$0.50 & 2 \\
329 Svea$^*$ & 2.477 & 3.9$\pm$0.5 & 69.2$\pm$0.7 & C & - & Y &4355$\pm$16 & 1.2$\pm$0.1 & 4185-4564 & 0.94$\pm$0.50 & 2 \\
331 Etheridgea & 3.024 & 4.5$\pm$0.3$^I$ & 74.9$\pm$2.7 & C  & C & Y & 6999$\pm$14 & 3.0$\pm$0.1 & 5556-8286 & 0.21$\pm$0.50 & 2 \\
334 Chicago & 3.895 & 5.1$\pm$1.6 & 174.1$\pm$12.8 & C & - & N & -& -& -& 0.66$\pm$0.50 & 2 \\
342 Endymion & 2.568 & 3.5$\pm$0.7 & 64.3$\pm$1.7 & C & Ch & Y & 6810$\pm$26 & 1.7$\pm$0.1 & 5356-8229 & -1.47$\pm$0.50 & 2 \\
356 Liguria$^+$ & 2.758 & 5.3$\pm$1.5 & 131.0$\pm$9.7 & C & - & N & -& -& -& -1.17$\pm$0.51 & 2 \\
360 Carlova$^*$ & 3.001 & 4.1$\pm$0.8 & 132.6$\pm$2.3 & C & C & Y & 4421$\pm$11 & 1.6$\pm$0.1 & 4271-4609 & 0.18$\pm$0.50 & 2 \\
372 Palma & 3.148 & 6.4$\pm$1.3 & 190.4$\pm$6.6 & C  & B & N & -& -& -& 0.33$\pm$0.50 & 2 \\
381 Myrrha & 3.224 & 5.3$\pm$1.4 & 129.0$\pm$11.6 & C & Cb & N & -& -& -& 2.43$\pm$0.50 & 2 \\
386 Siegena & 2.896 & 6.9$\pm$0.2$^I$ & 165.0$\pm$2.7 & C & C & Y & 6801$\pm$40 & 3.3$\pm$0.1 & 5419-8419 & 1.12$\pm$0.51 & 2 \\
393 Lampetia & 2.778 & 8.3$\pm$1.0$^I$ & 96.9$\pm$31.4 & C & Xc & Y & 7014$\pm$77 & 3.1$\pm$0.1 & 5427-8037 & 0.07$\pm$0.50 & 2 \\
395 Delia & 2.785 & 4.8$\pm$0.5$^I$ & 51.0$\pm$2.4 & C & Ch & Y & 6990$\pm$52 & 2.1$\pm$0.1 & 5538-8355 & 1.51$\pm$0.50 & 2 \\
405 Thia & 2.584 & 4.7$\pm$1.7 & 125.0$\pm$17.4 & C & Ch & Y & 6618$\pm$102 & 2.5$\pm$0.2 & 5523-8054 & -0.03$\pm$0.50 & 2 \\
407 Arachne & 2.625 & 5.5$\pm$0.7$^I$ & 95.1$\pm$5.4 & C & - & Y & 6912$\pm$80 & 2.8$\pm$0.1 & 5281-8345 & 0.67$\pm$0.50 & 2 \\
 & & & & & & & 8633$\pm$37 & 1.4$\pm$0.1 & 8203-8965 & & \\
409 Aspasia & 2.576 & 5.1$\pm$1.0 & 177.0$\pm$0.9 & C & Xc & N & -& -& -& 4.13$\pm$0.50 & 2 \\
410 Chloris & 2.725 & 4.3$\pm$0.7 & 118.9$\pm$2.9 & C & Ch & Y & 6952$\pm$25 & 3.1$\pm$0.3 & 5579-8292 & 1.58$\pm$0.51 & 1 \\
414 Liriope & 3.514 & 2.7$\pm$0.3 & 88.8$\pm$2.2 & C & Cg & Y & 6885$\pm$52 & 0.9$\pm$0.2 & 5409-8250 & -1.04$\pm$0.50 & 2 \\
419 Aurelia & 2.596 & 4.6$\pm$0.3$^I$ & 129.0$\pm$4.1 & F & - & N & -& -& -& 0.79$\pm$0.50 & 2 \\
444 Gyptis & 2.770 & 4.8$\pm$2.5 & 163.0$\pm$25.5 & C & C & Y & 4366$\pm$10 & 3.5$\pm$0.1 & 4193-4500 & 0.74$\pm$0.51 & 1 \\
 & & & & & & & 6909$\pm$33 & 1.5$\pm$0.1 & 5201-8497 & & \\
449 Hamburga & 2.551 & 3.9$\pm$0.2$^I$ & 85.6$\pm$1.9 & C & - & Y & 6941$\pm$20 & 2.3$\pm$0.1 & 5462-8419 & 0.40$\pm$0.50 & 2 \\
476 Hedwig & 2.650 & 4.9$\pm$0.2$^I$ & 116.8$\pm$2.6 & P & X & N & -& -& -& 3.40$\pm$0.50 & 2 \\
488 Kreusa & 3.167 & 5.9$\pm$2.2 & 150.0$\pm$11.3 & C & - & Y & 6692$\pm$44 & 3.4$\pm$0.1 & 5302-8398 & -0.76$\pm$0.50 & 1 \\
489 Comacina & 3.153 & 4.3$\pm$0.2$^I$ & 139.4$\pm$3.0 & C & - & N & -& -& -& 1.53$\pm$0.50 & 2 \\
490 Veritas & 3.171 & 6.2$\pm$0.6$^I$ & 115.6$\pm$5.5 & C & Ch & Y & 6675$\pm$10 & 6.7$\pm$0.1 & 5124-8455  & 0.18$\pm$0.54 & 1 \\
511 Davida & 3.163 & 7.2$\pm$1.3 & 283.3$\pm$4.0 & C & C & N & -& -& -& 3.03$\pm$0.50 & 2 \\
559 Nanon & 2.712 & 5.0$\pm$0.4$^I$ & 79.8$\pm$2.7 & C & Xk & N & -& -& -& 4.12$\pm$0.50 & 2 \\
583 Klotilde & 3.169 & 6.2$\pm$1.2 & 84.5$\pm$5.3 & C & - & Y & 6931$\pm$40 & 1.0$\pm$0.1 & 5573-8114 & -0.15$\pm$0.50 & 2 \\
585 Bilkis & 2.431 & 4.6$\pm$1.4 & 51.4$\pm$0.7 & C & - & Y & 7148$\pm$45 & 3.2$\pm$0.1 & 5521-8293 & -0.43$\pm$0.50 & 2 \\
602 Marianna & 3.090 & 5.2$\pm$0.7 & 126.8$\pm$2.1 & C & - & Y & 7003$\pm$57 & 3.4$\pm$0.1 & 5498-8248 & 0.13$\pm$0.50 & 2 \\
618 Elfriede & 3.191 & 5.1$\pm$0.6 & 131.2$\pm$1.1 & C & - & N & -& -& -& 1.66$\pm$0.50 & 1 \\
654 Zelinda & 2.297 & 4.3$\pm$1.1 & 127.0$\pm$20.5 & C & Ch & Y & 6724$\pm$24 & 2.3$\pm$0.1 & 5387-8389 & 0.74$\pm$0.50 & 2  \\		 
702 Alauda$^+$ & 3.193 & 5.5$\pm$1.1 & 202.0$\pm$4.6 & C & B & N & -& -& -& 0.15$\pm$0.50 & 2 \\
704 Interamnia & 3.063 & 7.6$\pm$2.8 & 312.0$\pm$39.8 & F & B & N &- &- &- & -0.18$\pm$0.50 & 2 \\
712 Boliviana & 2.575 & 3.9$\pm$0.5 & 127.6$\pm$3.7 & C & X & Y & 6744$\pm$73 & 1.6$\pm$0.1 & 5643-8601 & 0.62$\pm$0.50 & 1 \\
748 Simeisa & 3.961 & 4.2$\pm$0.2$^I$ & 103.0$\pm$2.2 & P & - & N & -& -&- & 4.95$\pm$0.50 & 2 \\
776 Berbericia & 2.932 & 6.6$\pm$0.8 & 151.1$\pm$4.1 & C & Cgh & Y & 6986$\pm$28 & 3.4$\pm$0.2 & 5420-8527 & -0.16$\pm$0.50 & 1 \\
786 Bredichina & 3.170 & 4.8$\pm$0.6 & 103.8$\pm$3.7 & C & - & N & -& -& -& 2.30$\pm$0.50 & 2 \\
790 Pretoria & 3.409 & 3.8$\pm$0.1$^I$ & 170.4$\pm$2.6 & P & - & N & -& -& -& 3.96$\pm$0.50 & 2 \\
814 Tauris & 3.152 & 4.4$\pm$0.7 & 109.9$\pm$1.9 & C & C & N & -& -& -& 1.32$\pm$0.50 & 2 \\
868 Lova & 2.704 & 5.5$\pm$0.7 & 51.2$\pm$0.6 & C & Ch & Y & 6907$\pm$47 & 1.0$\pm$0.1 & 5399-8322 & -0.43$\pm$0.50 & 2 \\
977 Philippa & 3.115 & 5.4$\pm$1.5 & 66.9$\pm$1.2 & C & - & Y & 6830$\pm$10 & 2.3$\pm$0.1 & 5483-8281 & 1.54$\pm$0.50 & 2 \\
1015 Christa & 3.207 & 4.6$\pm$0.6 & 96.6$\pm$2.9 & C & Xc & N & -& -& -& 0.65$\pm$0.50 & 2 \\
1021 Flammario & 2.737 & 4.7$\pm$2.0 & 98.0$\pm$8.4 & F & B & N & -& -& -& -0.38$\pm$0.50 & 2 \\
1093 Freda & 3.130 & 3.8$\pm$0.2$^I$ & 116.7$\pm$2.9 & C & - & N & -& -& -& 2.13$\pm$0.51 & 1 \\
1268 Libya & 3.983 & 4.5$\pm$0.2$^I$ & 94.1$\pm$2.3 & P & - & N & -& -& -& 3.16$\pm$0.51 & 2 \\
1330 Spiridonia & 3.170 & 3.1$\pm$0.5 & 69.9$\pm$1.1 & P & - & N &- &- &- & 1.32$\pm$0.50 & 2 \\
\hline 
\end{longtable}
\end{center}
}
\begin{list}{}{}
\item 
\end{list}


{\small
\begin{table}
\begin{center}
\caption{Comparison between the 0.7 $\mu$m and the 3 $\mu$m band data. For the 3 $\mu$m band, keys to reference are the following: (1) Lebofsky et al., 1990; (2) Jones et al., 1990; (3) Takir \& Emery, 2012; (4) Feierberg et al., 1985; (5) Rivkin, 1997; (6) Rivkin \& Emery, 2010; (7) Lebofsky, 1980; (8) Campins et al., 2010; (9) Licandro et al., 2011; (10) Howell, 2014, personal communication; (11) Howell et al., 2001 ; (12) Howell et al., 2011.} 
\label{tab3micron}
\begin{tabular}{|l|c|l|l|c|l|} \hline
{\bf ASTEROID} & {\bf 0.7 \boldmath{$\mu$}m} & {\bf 3 \boldmath{$\mu$}m band}&  {\bf ASTEROID} & {\bf 0.7 \boldmath{$\mu$}m} & {\bf 3 \boldmath{$\mu$}m band}  \\ \hline
10 Hygiea & Y & Wet$^{1,2,3,4}$ & 13 Egeria & Y & Wet$^{2,4,5}$   \\
19 Fortuna & Y & Wet$^{2,4,5,7}$   & 24 Themis & Y & Wet$^{1}$/H$_2$0$^{3,6,8}$   \\ 
 34 Circe & Y & Wet$^{3}$    & 36 Atalante & Y & Wet$^{3,4}$     \\
 48 Doris & Y & Wet$^{3}$    & 51 Nemausa & Y & Wet$^{1,2,5,7}$    \\ 
 54 Alexandra & Y & Wet$^{3}$    & 65 Cybele & N & Dry$^{1,4,7}$/ H$_2$O$^{9}$   \\
 70 Panopaea & Y & Wet$^{7,5}$    & 74 Galatea & N & Dry$^{4}$  \\
 104 Klymene & Y & Wet$^{3}$    & 107 Camilla & N & H$_2$0$^{3}$  \\
 120 Lachesis & N &Wet$^{3}$    & 121 Hermione & Y & Wet$^{3}$ \\ 
130 Elektra & Y & Wet$^{2,3}$   & 140 Siwa & N & Dry$^{3}$ \\
153 Hilda & N & H$_2$0$^{3}$ & 185 Eunike & N & Dry$^{5}$   \\
190 Ismene & N & H$_2$0$^{3}$/Dry$^{10}$   & 240 Vanadis & Y & Wet$^{5}$   \\
304 Olga & N & Dry$^{5}$   & 313 Chaldea & Y & Wet$^{2}$    \\
334 Chicago & N & Wet$^{3}$    & 386 Siegena & Y & Wet$^{4}$   \\
409 Aspasia & N & Wet$^{2}$/Dry$^{5}$?  & 410 Chloris & Y & Wet$^{4}$   \\
444 Gyptis & Y & Dry$^{5}$   & 449 Hamburga & Y & Wet$^{10}$/Dry$^{11}$   \\
476 Hedwig & N & Wet$^{12}$   & 511 Davida & N & Wet$^{2,3,4}$   \\
 704 Interamnia & N & Wet$^{1,3}$/Dry?$^{2,4}$  & 748 Simeisa & N & Dry$^{1}$ \\
 776 Berbericia & Y & Wet$^{4}$  & 790 Pretoria & N & H$_2$0$^{3}$  \\
 1015 Christa & N & Wet$^{3}$ & & &   \\
\hline
\end{tabular}
\end{center} 
\end{table}
}

\newpage

{\small
\begin{table}
\begin{center}
\caption{Number of the observed asteroids (N$_{TOT}$) for each taxonomic class. The number of hydrated and non hydrated asteroids (N$_{HY}$ and N$_{ANHY}$, respectively) and the associated mean albedo values are reported (for the albedo, we specify in parenthesis the number of asteroids for which this value is available).
} 
\label{summary}
\begin{tabular}{|l|c|c|c|c|c|c|c|} \hline
 Class & N$_{TOT}$ & N$_{HY}$ & N$_{ANHY}$ &  Perc$_{HY}$ (\%) & albedo$_{HY}$ (\%) & albedo$_{ANHY}$ (\%) \\ \hline \hline
C &    454 &  230 &  224 & 50.7 & 5.7$\pm$2.1 (219) & 6.1$\pm$2.6 (204) \\
F &     13 &    1 &   12 &  7.7 & 6.3$\pm$2.4 (1) & 5.2$\pm$1.9 (12) \\
B &     92 &    9 &   83 &  9.8 & 7.5$\pm$2.5 (8)    &  8.0$\pm$3.2 (76) \\
G &     18 &   18 &   0 & 100  & 8.8$\pm$3.8  (16) & -- \\
P &     23 &    1 &  22 &  4.3 &  7.1$\pm$0.3   (1) & 5.0$\pm$2.5 (21) \\ \hline
\end{tabular}
\end{center} 
\end{table}
}

 
{\scriptsize
\begin{center}
\begin{longtable}{|l|c|c|c|c|c|c|c|c|c|c|c|}
\caption[]{Identification of the 0.7 $\mu$m absorption band on the primitive asteroids found in the literature.
\textbf{Notes.} We applied exactly the same analysis as done for the asteroids presented in Table 2.
The albedo comes from WISE data when available, or from IRAS (indicated with the symbol $^I$), and in few cases from 
the MSX Infrared MPS (symbol $^M$).
When the Tholen taxonomic class is complex (several possible groups) or doesn't exist, we choose the class based on the available spectra. 
The references in bold font are the measures kept for the
statistics. When there is no bold font, please refer to Table 2.
\textbf{References}: (1) Fornasier et al., 2011; (2) Xu et al., 1995 (SMASS survey); (3) Bus \& Binzel, 2002 (SMASS II survey); (4)  Lazzaro et al., 2004 (S3OS2 survey); (5) Sawyer, 1998 ($^{a}$: files 1 and 2;
$^{b}$: file 4; $^{c}$: file 5; $^{d}$: file 6; $^{e}$: file 3; $^{f}$:
file 2; $^{g}$: file 1; $^{h}$: files 1 and 3.; (6) Vilas, 1998.}
\label{tab_all} \\
\hline \multicolumn{1}{|c|} {\textbf{Asteroid}} & \multicolumn{1}{c|}
{\textbf{a}} &\multicolumn{1}{c|}
{\textbf{Alb.}} & \multicolumn{1}{c|} {\textbf{Diam.}} & \multicolumn{2}{c|}
{\textbf{Taxon.}} & \multicolumn{1}{c|} {\textbf{H}} &     \multicolumn{1}{c|} {\textbf{Center}} & \multicolumn{1}{c|} {\textbf{Depth}} & \multicolumn{1}{c|}{\textbf{Width}} & \multicolumn{1}{c|} {\textbf{Slope}} & \multicolumn{1}{c|}{\textbf{R}}\\  
\multicolumn{1}{|c|}{\textbf{}}  & \multicolumn{1}{c|}{\textbf{(AU)}}
 & \multicolumn{1}{c|} {\textbf{(\%)}} & \multicolumn{1}{c|}{\textbf{(km)}}
 & \multicolumn{1}{c|} {\textbf{(Th)}} &     \multicolumn{1}{c|}{\textbf{(B)}}  & \multicolumn{1}{c|}{\textbf{}} & \multicolumn{1}{c|}{\textbf{(\AA)}} & \multicolumn{1}{c|}{\textbf{(\%)}} & \multicolumn{1}{c|}{\textbf{(\AA)}} & \multicolumn{1}{c|}{\textbf{\%/($10^{3}$\AA)}} & \\ \hline
\endfirsthead
\multicolumn{12}{c}%
{{\bfseries \tablename\ \thetable{} -- continued from previous page}} \\ \hline 
\endfoot
\hline 
\hline \multicolumn{1}{|c|} {\textbf{Asteroid}} & \multicolumn{1}{c|}
{\textbf{a}} &\multicolumn{1}{c|}
{\textbf{Alb.}} & \multicolumn{1}{c|} {\textbf{Diam.}} & \multicolumn{2}{c|}
{\textbf{Taxon.}} & \multicolumn{1}{c|} {\textbf{H}} &     \multicolumn{1}{c|} {\textbf{Center}} & \multicolumn{1}{c|} {\textbf{Depth}} & \multicolumn{1}{c|}{\textbf{Width}} & \multicolumn{1}{c|} {\textbf{Slope}} & \multicolumn{1}{c|}{\textbf{R}}\\  
\multicolumn{1}{|c|}{\textbf{}}  & \multicolumn{1}{c|}{\textbf{(AU)}}
 & \multicolumn{1}{c|} {\textbf{(\%)}} & \multicolumn{1}{c|}{\textbf{(km)}}
 & \multicolumn{1}{c|} {\textbf{(Th)}} &     \multicolumn{1}{c|}{\textbf{(B)}}  & \multicolumn{1}{c|}{\textbf{}} & \multicolumn{1}{c|}{\textbf{(\AA)}} & \multicolumn{1}{c|}{\textbf{(\%)}} & \multicolumn{1}{c|}{\textbf{(\AA)}} & \multicolumn{1}{c|}{\textbf{\%/($10^{3}$\AA)}} & \\ \hline
\hline
\endhead
\hline \multicolumn{12}{r}{{Continued on next page}} \\ 
\endfoot
\hline \hline
\endlastfoot
1 Ceres & 2.768 & 11.3$\pm$0.5$^I$& 848.4$\pm$19.7 & G & C &  N  &   -   &   -   &   -   &   -   & 3,4,5 \\
2 Pallas  &  2.772  &  14.2$\pm$5.0  &   544.0$\pm$74.4  &    B  &    B  &   N  &   -   &   -   &   -   &    0.58$\pm$0.09  & 3,$5^a$,\textbf{6} \\
10 Hygiea & 3.137 & 5.8$\pm$0.5 & 453.2$\pm$19.2 & C & C & N & - &-& - &- & 2,3 \\
 & & & & & & Y & 6702$\pm$72 & 3.3$\pm$0.1& 2655 & 0.22$\pm$0.11 & 5 \\
13 Egeria & 2.576 & 6.9$\pm$2.2 & 227.0$\pm$25.9 & G & Ch & Y &  6933$\pm$38  &  2.7$\pm$0.1  &  2888  &   -0.09$\pm$0.13  & 3 \\
 & & & & & & Y & 6903$\pm$13  &  6.2$\pm$0.1  &  2905  &   -0.23$\pm$0.18 & 5 \\
19 Fortuna & 2.442 & 5.0$\pm$2.0 & 223.0$\pm$43.6 & G & Ch & Y &  6908$\pm$14  &  2.5$\pm$0.1  &  2775  &   -0.21$\pm$0.13  & 3 \\
 & & & & & & Y & 6987$\pm$27  &  4.5$\pm$0.1  &  2990  &   -0.73$\pm$0.13 & 5 \\
\hline 
\end{longtable}
\end{center}
}
\begin{list}{}{}
\item 
\end{list}


\newpage

{\bf Figures}

\end{document}